\documentclass[5p,times]{elsarticle}

\usepackage[hyphens]{url}
\usepackage{hyperref}
\usepackage{adjustbox}
\usepackage{booktabs}
\usepackage{color}
\usepackage{listings}
\usepackage{makecell}

\definecolor{codeblue}{rgb}{0,0,0.75}
\definecolor{codegreen}{rgb}{0,0.5,0}
\definecolor{codepurple}{rgb}{0.58,0,0.82}
\definecolor{codegray}{rgb}{0.5,0.5,0.5}

\lstdefinestyle{python_style}{
    commentstyle=\color{codegreen},
    keywordstyle=\color{codeblue},
    numberstyle=\tiny\color{codegray},
    stringstyle=\color{codepurple},
    basicstyle=\ttfamily\scriptsize,
    breakatwhitespace=false,
    breaklines=true,
    captionpos=b,
    keepspaces=true,
    showspaces=false,
    showstringspaces=false,
    showtabs=false,
	frame=tb
}

\graphicspath{{figures/}}

\journal{Future Generation Computer Systems}

\begin{document}

\clearpage
\thispagestyle{empty}
\noindent\fbox{%
	\parbox{\textwidth}{%
	\vspace{10pt}\large \centering CC-BY 4.0. This is the author's preprint version of the article ``Triggerflow: Trigger-based Orchestration of Serverless Workflows'' published in journal Future Generation Computer Systems (Volume 124, Pages 215-229).\\\texttt{DOI: 10.1016/j.future.2021.06.004}\vspace{10pt}
	}%
}
\clearpage
\pagenumbering{arabic}

\begin{frontmatter}

\title{Triggerflow: Trigger-based Orchestration of Serverless Workflows}

\author[urv]{Aitor Arjona}
\ead{aitor.arjona@urv.cat}
\author[urv]{Pedro~Garc\'ia L\'opez}
\ead{pedro.garcia@urv.cat}
\author[urv]{Josep~Samp\'e}
\ead{josep.sampe@urv.cat}
\author[ibm]{Aleksander Slominski}
\ead{aslom@us.ibm.com}
\author[ibm]{Lionel Villard}
\ead{villard@us.ibm.com}

\address[urv]{Universitat Rovira i Virgili, Tarragona, Spain}
\address[ibm]{IBM Watson Research, New York, USA}

\begin{abstract}
As more applications are being moved to the Cloud thanks to serverless computing, it is increasingly necessary to support the native life cycle execution of those applications in the data center. 

But existing cloud orchestration systems either focus on short-running workflows (like IBM Composer or Amazon Step Functions Express Workflows) or impose considerable overheads for synchronizing massively parallel jobs (Azure Durable Functions, Amazon Step Functions). None of them are open systems enabling extensible interception and optimization of custom workflows.

We present Triggerflow: an extensible Trigger-based Orchestration architecture for serverless workflows. We demonstrate that Triggerflow is a novel serverless building block capable of constructing different reactive orchestrators (State Machines, Directed Acyclic Graphs, Workflow as code, Federated Learning orchestrator). We also validate that it can support high-volume event processing workloads, auto-scale on demand with scale down to zero when not used, and transparently guarantee fault tolerance and efficient resource usage when orchestrating long running scientific workflows.
\end{abstract}


\end{frontmatter}

%\linenumbers

\section{Introduction}

Serverless Function as a Service (FaaS) is becoming a very popular programming model in the cloud thanks to its simplicity, billing model and inherent elasticity. The FaaS programming model is considered event-based, since functions are activated (triggered) in response to specific Cloud Events (like a state change in a disaggregated object store like Amazon S3).

The FaaS model has also proven ideally suited (PyWren \cite{pywren}, ExCamera\cite{fouladi}) for executing embarrassingly parallel computing tasks. But both PyWren and ExCamera required their own ad-hoc external orchestration services to synchronize the parallel executions of functions. For example, when the PyWren client launches a map job with N functions, it waits and polls Amazon S3 until all the results are received in the S3 bucket. ExCamera also relied on an external Rendezvous server to synchronize the parallel executions.

%And more recently, a plethora of academic work is flourishing to propose optimized schedulers and DAG (Direct Acyclic Graph) Engines over serverless functions \cite{ibmLithops, wukong, ripple, predictive, malawski, formal} to execute scientific workflows. The problem of all those academic works is that they rely on ad-hoc serverful schedulers which are themselves not serverless. This means that they are not designed as cloud native managed services guaranteeing scalability, flexible scaling, fault tolerance or pay as you go models.  

Lambda creator Tim Wagner recently outlined \cite{wagner} that Cloud providers must offer new serverless building blocks to applications. In particular, he foresees new services like fine-grained, low-latency orchestration, execution data flows, and the ability to customize code and data at scale to support the emerging data-intensive applications over Serverless Functions.

The reality is that existing serverless orchestration systems are not designed for long-running data analytics tasks \cite {wosc4,wosc5}. Either they are focused on short-running highly interactive workflows (Amazon Express Workflows, IBM Composer) or impose considerable overheads for synchronizing massively parallel jobs (Azure Durable Functions, Amazon Step Functions, Google Cloud Workflows). %None of them are open systems enabling extensible interception and optimization of custom workflows.

We present Triggerflow, a novel building block for composing event-based services. As more applications are moved to the Cloud, this service will enable to control the life-cycle of those applications in a reactive and extensible way. The flexibility of the system can also be used to transparently optimize the execution of tasks in reaction to events.

%. Our proposal is not just another scheduler or specialized engine for serverless functions, but a generic multi-tenant service designed for the creation of different reactive  scheduling services in the Cloud. 

% We claim that if serverless functions follow a trigger-based model, the serverless orchestration system should also be trigger-based. This means that in a DAG (direct acyclic graph) workflow, the termination of one or many functions should trigger the next stage using asynchronous events.

%If the orchestration system is not trigger-based, reactive and asynchronous, the orchestrator will require some synchronous blocking actions to wait the termination of functions, or the transition from one stage to the following. This also implies an active orchestrator with increased billing for long-running workflows.

The major contributions of this paper are the following:
\begin{enumerate}

\item We present a Rich Trigger framework following an Event-Condition-Action (ECA) architecture that is extensible at all levels (Event Sources and Programmable Conditions and Actions). Our architecture ensures that composite event detection and event routing mechanisms are mediated by reactive event-based middleware.

\item We demonstrate Triggerflow's extensibility and universality creating atop it a state machine workflow orchestrator, a DAG engine, an imperative Workflow as Code (using event sourcing) orchestrator, integration with an external scheduler like Lithops \cite{lithops} and a Federated Learning orchestrator. We also validate performance and overhead of our orchestration solution compared to existing Cloud Serverless Orchestration systems like Amazon Step Functions, Amazon Express Workflows, Azure Durable Functions and IBM Composer.

\item We demonstrate how Triggerflow is reactive and scales on demand, using an event-based autoscaler component that provisions resources to the system only when events are produced. With scale to zero, Triggerflow follows a serverless-like pay-per-use model, making an efficient use of compute resources.

\item We finally propose a generic implementation of our model over standard CNCF or Open Source production-grade technologies like Kubernetes, KEDA, Knative and CloudEvents. We validate that our system can support high-volume event processing workloads, auto-scale on demand and transparently optimize scientific workflows. The project is available as open-source in \cite{triggerflow}.
%We also analyze the feasibility of a multi-tenant implementation of this service and major deployment challenges for the future. 

\end{enumerate}

\section{Related work}

FaaS is based on the event-driven programming model. In fact, many event-driven abstractions like triggers, Event Condition Action (ECA) and even composite event detection were already inspired by the veteran Active Database Systems \cite{active}. 

%And event-driven systems have become the backbone of a myriad of distributed systems because of their decoupled and scalable architecture. The reactive manifesto \cite{rmanifesto} outlines four interesting properties of such systems like their elasticity, resiliency, responsiveness and message-driven (event-driven) asynchronous communication model ensuring loose coupling.

Event-based triggering has also been extensively employed in the past to provide reactive coordination of distributed systems \cite{mitchell2012oolong, han2013large}. Event-based mechanisms and triggers have also been extensively used \cite{eve, chen2008,binder2006, padres} in the past to build workflows and orchestration systems. The ECA model including trigger and rules fits nicely to define the transitions of finite state machines representing workflows. In \cite{dai2018trigger}, they propose to use synchronous aggregation triggers to coordinate massively parallel data processing jobs.

An interesting related work is \cite{padres}. They leverage composite subscriptions in content-based publish/subscribe systems to provide decentralized Event-based Workflow Management. Their PADRES system supports parallelization, alternation, sequence, and repetition compositions thanks to content-based subscriptions in a Composite Subscription Language. %They propose a decentralized orchestration service where job agents subscribe to the previous stage in a distributed fashion. 

More recently, a relevant article \cite{cep} has surveyed the intersections of the Complex Event Processing (CEP) and Business Process Management (BPM) communities. They clearly present the existing challenges to combine both models and describe recent efforts in this area. We outline that our paper is in line with their challenge ``Executing business processes via CEP rules", and our novelty here is our serverless reactive and extensible architecture.

In serverless settings, the more relevant related work aiming to provide reactive orchestration of serverless functions is the Serverless trilemma \cite{trilemma} from IBM. In their paper, the authors advocate for reactive run-time support for function orchestration, and present a solution for sequential compositions on top of Apache OpenWhisk.

Recently, effort from the CNCF community has been put into creating a standard specification for Serverless Workflows \cite{serverlessworkflow}. They propose a declarative definition of a workflow as a YAML file that contains descriptions for CloudEvents to consume, event-driven invocation of serverless functions and state transitions for workflow data management and control flow logic. The idea is to define an abstract definition that can be interpreted by different systems thus ensuring portability and to avoid vendor lock-in.

%An interesting related work is the gg \cite{gg} framework for building burst-parallel cloud-functions application. gg leverages existing stateless applications like software compilation, unit tests,  or  video encoding and deployes them using a composition of lightweight OS containers, taking care of dependencies, data movements, and dealing with failures and stragglers. Nevertheless, gg does not follow a event-based architecture nor it provides an extensible layer for building different schedulers.

A plethora of academic works are proposing different so-called serverless orchestration systems like \cite{wukong, ripple,  malawski, formal, specrg, gg}. However, most of them rely on centralized serverful components like VMs or dedicated resources that do not scale down to zero. Instead, the orchestrator component is active during the whole workflow execution. This results in inefficient resource usage for long-running workflows because the orchestrator will stand idle most of the time waiting for long tasks to finish. Other use functions calling functions patterns which complicate their architectures and fault tolerance. None of them offer extensible trigger abstractions to build different orchestrators.

Another related work is \cite{burckhardt2021serverless}. The authors compare Durable Functions (workflow as code) and triggers for workflow orchestration. They claim that using triggers is possible for workflow orchestration but that it is not ideal. The main drawbacks are that (i) it is necessary to create different queues/directories for each step, (ii) triggers cannot wait for the completion of multiple previous steps, and (iii) triggers are not suitable for correct error handling. This is true for conventional triggers. However, in this article we will see that using a Rich Trigger framework can resolve these problems. With extended trigger logic we can specify rules to filter events (to avoid creating multiple queues) and to aggregate events (to perform a multiple join). With event replay and checkpointing we can also guarantee fault tolerance (Section \ref{sec:fault_tolerance}). In fact, we demonstrate how using dynamic and flexible triggers we can orchestrate workflows defined as code (Section \ref{sec:workflow_as_code}).

All Cloud providers are now offering cloud orchestration and function composition services like IBM Composer, Amazon Step Functions, Azure Durable Functions, or Google Cloud Workflows.

IBM Composer service is in principle designed for short-running synchronous composition of serverless functions. IBM Composer generates a state machine representation of the workflow to be executed with IBM Cloud Functions. It can represent sequences, conditional branching, loops, parallel, and map tasks. However, fork/join synchronization (map, parallel) blocks on an external user-provided Redis service, limiting their applicability to short running tasks.

%that generates a state machine representation of the workflow  to be executed with IBM Cloud Functions.  Their library can represent sequences, conditional branching, loops, parallel, and map tasks.

%although it requires the user to provide an external Redis service (for synchronizing fork/join maps and parallels). 

Amazon offers two main services: Amazon Step Functions (ASF) and Amazon Step Functions Express Workflows (ASFE). The Amazon States Language (based on JSON) permits to model task transitions, choices, waits, parallel, and maps in a standard way. ASF is a fault-tolerant managed service designed to support long-running workflows and ASFE is designed for short-running (less than five minutes) highly intensive workloads with relaxed fault-tolerance.
%Both rely on a declarative state machine language that enables the composition of serverless functions (Lambda) but also of other resources in the AWS Cloud like EC instances and VMs. 

Microsoft's Azure Durable Functions (ADF) represents workflows as code using C\# or Javascript,  leveraging async/await constructs and using event sourcing to replay workflows that have been suspended.  ADF does not support map jobs explicitly, and only includes a \emph{Task.whenAll} abstraction enabling fork/join patterns for a group of asynchronous tasks.
 %ADF features a rich but advanced programming model even including a Durable stateful entity resembling an actor.
 
%Google offers Google Cloud Composer service leveraging a managed Apache Airflow cluster.  Airflow represents workflows in a DAG (Directed Acyclic Graph) coded in Python, so that it cannot support cycles. It is not ideally suited for parallel jobs or high-volume workflows, and it is not designed for orchestrating serverless functions.

Google Cloud offers Google Cloud Workflows service. Workflows in Google Cloud Workflows are represented as a series of steps with basic logical flow control like conditions or loops. Every step makes an HTTP request that can be used, for example, to trigger a Google Cloud Function. It is not designed for broad parallel tasks as it lacks the \textit{map} primitive present in other systems like ASF.

Two previous papers \cite{wosc4,wosc5} have compared public FaaS orchestration services for coordinating massively parallel workloads.  In those studies, IBM Composer offered the fastest performance and reduced overheads to execute map jobs whereas ASF or ADF imposed considerable overheads.  We will also show in this paper how ASFE obtains good performance for parallel workloads.

%But both IBM Composer and ASFE are designed for short-running workflows, which compromise their applicability for long-running scientific workflows.  ASFE is limited to five minutes workflows, and it relaxes fault tolerance support compared to ASF. IBM Composer is also designed for short-running synchronous tasks, in particular the blocking wait for map jobs using an external user-provided Redis service. 
% IBM Composer is not providing guarantees for workflow exection. On the contrary, the reactive event-based  architecture of TriggerFlow supports long-running high volume workflows with good performance and guaranteed message processing as we demonstrate in the validation. 

%Our approach advocates that orchestration must be trigger-based, and offered by an external scalable  service decoupled from the FaaS runtime due to scalability and fault-tolerance reasons. At least two existing systems (OpenWhisk, Fission) and two academic papers\cite{formal, specrg}  propose designs where orchestration is inter-twinned with the FaaS run-time architecture.  

None of the existing cloud orchestration services is offering an open and extensible trigger-based API enabling the creation of custom workflow engines. We demonstrate in this paper that we can use Triggerflow to implement existing models like ASF or Airflow DAGs. Triggerflow is not just another scheduler, but a reactive meta-tool to build reactive orchestrators leveraging Kubernetes standard technologies.

%And they do not offer transparent interception mechanisms to optimize existing workflows.

\subsection{Cloud Event Routing and Knative Eventing}

Event-based architectures are gaining relevance in Cloud providers as a unifying infrastructure for heterogeneous cloud services and applications. Event services participate in the entire cloud control loop from event production in event sources, to event detection using monitoring services, to event logging and data analytics of existing event workflows, and finally to service orchestration and event reaction thanks to appropriate filtering mechanisms.

The trend is to create cloud event routers, specialized rule-based multi-tenant services, capable of filtering and triggering selected targets in the Cloud in response to events. Amazon is offering EventBridge, Azure offers EventGrid, and Google and IBM are investing in the open Knative Eventing project and CNCF CloudEvents standard.

%More recently, Amazon has introduced AWS Event Bridge, a generic event router permitting the creation of third-party event sources and making them trigger cloud native services (Targets). For example, AWS support targets such as Lambda functions, EC2 instances, AWS Step Functions state machines, streams in Kinesis, SNS topics, SQS queues and many others. EventBridge is another clear step to move filtering and event routing from Lambda functions to specialized rule-based multi-tenant services

%It provides all necessary building blocks to build highly scalable applications but it does not have an opinionated way of how to build cloud ``native" applications based on the best practices.
%Many of existing open-source FaaS frameworks are now targeting Kubernetes (Fission\cite{fission}, Kubeless\cite{kubeless}, OpenFaas\cite{openfaas}).

The Knative project was created to provide stream-lined serverless-like experience for developers using Kubernetes. It contains a set of high-level abstractions related to scalable functions (Knative Serving) and event processing (Knative Eventing) that allows the description of asynchronous, decoupled, event-driven applications built out of event sources, sinks, channels, brokers, triggers, filters, sequences, etc.

%Many of existing open-source FaaS frameworks are now targeting Kubernetes (Fission\cite{fission}, Kubeless\cite{kubeless}, OpenFaas\cite{openfaas}). Kubernetes is an open-source container orchestration platform that is very flexible and extensible.  Knative project \cite{knative} was created to provide streamlined serverless-like experience for developers using Kubernetes. It provides a set of higher level abstractions related to scalable functions (Knative Serving) and events processing that allows the description of asynchronous, decoupled, event-driven applications built out of event sources, sinks, channels, brokers, triggers, filters, sequences etc. (Knative Eventing). 

The goal of Knative is to allow developers to build cloud native event-driven serverless applications on those abstractions. The value of Knative is to encapsulate well tested best practices in high-level abstractions that are native to Kubernetes: custom resource definitions (CRDs) for new custom resources (CRs) such as event sources. Abstractions allow developers to describe event-driven application components and have late-binding to underlying (possibly multiple) messaging and eventing systems like Apache Kafka and NATS among others.

%One of the hardest problems in event-driven applications is to deal with reliability and scalability. Event systems may be receiving events as soon as they are created ("pushed") or they may process them when they are ready ("pull" or "poll") and for both cases they need to deal with capacity limits and error handling. Knative is very well suited for push-based scaling as it can autoscale based on incoming HTTP requests containing events. For pull-based scaling, Kubernetes Event-driven Autoscaling (KEDA)\cite{keda} can be used. Developers declare in CRs description of their application components including autoscaling, retrying, and error handling (for example using "dead letter channels" to declare where undelivered events should go) and leave implementation details to the underlying event routing system. 

%Currently Knative is before 1.0 release. There are several areas identified as gaps in Knative Eventing such as improvements to performance and scalability  and convergence with the KEDA \cite{keda} project. 

Triggerflow aims to leverage existing event routing technology (Knative Eventing) to enable extensible trigger-based orchestration of serverless workflows. Triggerflow includes advanced abstractions not present in Knative Eventing like dynamic triggers, trigger interception, custom filters, termination events, and a shared context among others. Some of these novel services may be adopted in the future by event routing services to make it easier to compose, stream, and orchestrate tasks.

\section{Triggerflow Architecture}

\medskip
\begin{figure*}[t]
	\includegraphics[width=\textwidth]{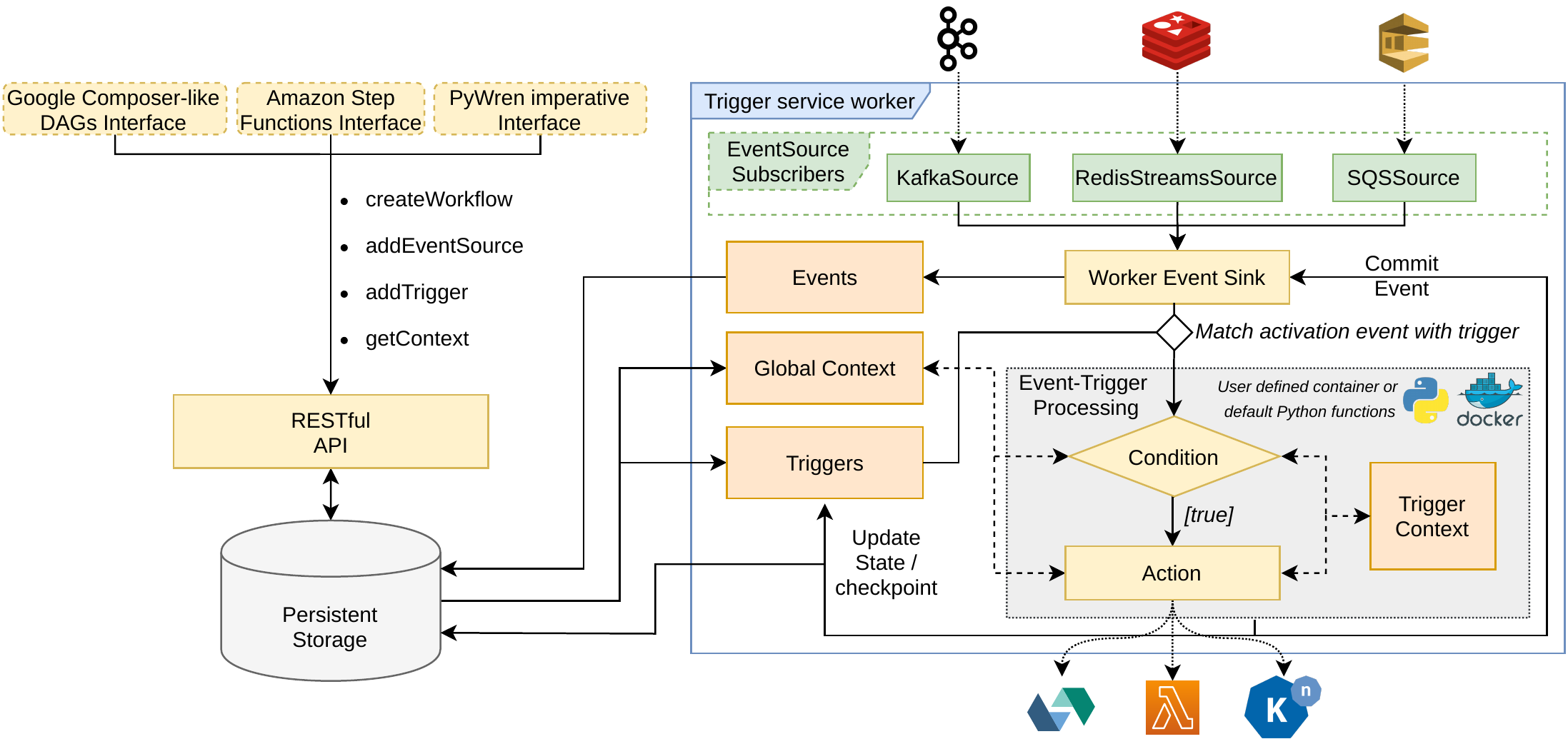}
	\centering
	\caption{Triggerflow Architecture}
	\label{fig:Triggerflow_architecture}
	\vspace{-8pt}
\end{figure*}

We can see in Figure~\ref{fig:Triggerflow_architecture} an overall diagram of the Triggerflow Architecture. The Trigger service follows an extensible Event-Condition-Action architecture. The service can receive events from different Event Sources in the Cloud (Kafka, RabbitMQ, Object Storage, timers). It can execute different types of Actions (containers, functions, VMs), and it can also enable the creation of custom filters or Conditions from third-parties. The Trigger service also provides a shared persistent context repository providing durability and fault tolerance.

We define Triggerflow as a Rich Trigger framework. A Rich Trigger framework differs from a regular triggering framework in that the former contains built-in programmable abstractions for extended event processing logic like composite event detection, event aggregation, event routing or stateful event processing and filtering, all with transparent fault tolerance.

Figure~\ref{fig:Triggerflow_architecture} also shows the basic API exposed by Triggerflow: \textit{createWorkflow} initializes the context for a given workflow, \textit{addTrigger} adds a new trigger (including event, conditions, actions, and context), \textit{addEventSource} permits the creation of new event sources, and \textit{getState} obtains the current state associated to a given trigger or workflow.

Different applications and orchestrators can benefit from serverless awakening and rich triggering by using this API to build different orchestration services like Airflow-like DAGs, ASF state machines or Workflow as Code clients like Lithops \cite{lithops}.

\subsection{Design goals}

Let's establish a number of design goals that must be supported in the proposed architecture:

\begin{enumerate}

\item Support for Heterogeneous Workflows: The main idea is to build a generic building block for different types of orchestrators. The system should support enterprise workflows based on Finite State Machines, Directed Acyclic Graphs, and Workflow as Code systems.  %The system should also allow to build external third-party schedulers that need to be awakened by state transitions.  %Finally, the system must support abstractions like branching, conditionals, iterations, parallels, maps and nested workflow executions.

\item Extensibility and Computational Reflection: The system must be extensible enough to support the creation of novel workflow systems with special requirements like specialized scientific workflows. The system must support introspection and interception mechanisms enabling the monitoring and optimization of existing workflows.

\item Serverless design: The system must be reactive, and only execute logic in response to events, like state transitions. Serverless design also entails pay per use, flexible scaling, and dependability. %This permits  pay per use, or billing the user only for the  resources that it is using.  The system must also support flexible scaling and dependability  for heterogeneous workloads.  

\item Performance: The system should support high-volume workloads like data analytics pipelines with numerous parallel tasks. The system should exhibit low overheads for both short-running and long-running workflows. 

\end{enumerate}

\subsection{Trigger service}

Our proposal is to design a purely event-driven and reactive architecture for workflow orchestration. Like previous works \cite{eve, chen2008, binder2006}, we propose to handle state transitions using event-based triggering mechanisms. The novelty of our approach precisely relies on the aforementioned design goals: support for heterogeneous workflows, extensibility, serverless design, and performance for high volume workloads.

We follow an \textbf{Event Condition Action} architecture in which triggers (active rules) define which action must be launched in response to Events or to Conditions evaluated over one or more Events. The system must be extensible at all levels: Events, Conditions, and Actions. Let us introduce some definitions:
% for our event-based orchestration model. 

\textbf{Definition 1. Workflow:} We can represent a workflow as a Finite State Machine (FSM) being a 6-tuple with 
\newline
 M = ($\sum_{in},Ctx, S, s,F, \delta$), 
in this 6-tuple:
\begin{enumerate} 
\item  $\sum_{in}$: the set of input events
\item  Ctx: the set of context variables
\item  S: the set of states which map to Actions in the ECA model
\item  s: initial state, linked to an initial event
\item  F: end state, linked to a final Termination event
\item  $\delta$: state-transition function: $\delta:  S \times \sum  \rightarrow S $ , based on the ECA triggers
\end{enumerate}

\textbf{Definition 2. Trigger ($\delta$):} can be defined as the state transition function. The trigger is a 4-tuple with (Event, Context, Condition, Action) that moves one state to the following when the condition on input events holds. In this case, the trigger launches the appropriate action which corresponds to the next state. Each action will in turn fire events that may be captured by another trigger. Triggers can be transient and dynamic (activated on demand) or persistent if they remain always active. 

Its components are:

\begin{itemize}
  \item \textbf{Event}: Events are the atomic piece of information that drive flows in Cloud applications. We rely on the standard CNCF CloudEvents version 1.0 specification to represent events. To match an event to its trigger, the \textit{subject} and \textit{type} fields of a CloudEvent are used.  We use the \textit{subject} field to match the event to its corresponding trigger, and the \textit{type} field to describe the type of the event. Termination and failure events use this \textit{type} field to notify success (and result) or failure (and code or error information).
 
 % To match an event to its trigger, the \textit{subject} and \textit{type} fields of a CloudEvent are used.  %According to CloudEvents version 1.0 specification \cite{cloudevents}, the \textit{subject} field is an optional parameter used to facilitate the filtering on generic subscription filtering scenarios. The \textit{type} field is required and it's used to describe the type of the event. For example, it can be used to determine whether a serverless function has been successful or has failed its execution, allowing the user to create special triggers for error handling.
  \item \textbf{Context}: The context is a fault-tolerant key-value data structure that contains the state of the trigger during its lifetime. It is also used to introspect the current trigger deployment, to modify the state of other triggers or to dynamically activate/deactivate triggers. % It is used, for example, to store a counter to join multiple serverless functions.
  \item \textbf{Condition}: Conditions are active rules (user-defined code) that filter events to decide if they match in order to launch the corresponding action. Conditions evaluate rules over primitive  events (single) or over composite (group) events. Composite event information like counters may be stored in the Context. Conditions produce a \textit{boolean} result that represents whether the trigger has to be fired or not.
  \item \textbf{Action}: Actions are the computations (user-defined code) launched in response to matching Conditions in a trigger. An Action can be used to asynchronously invoke a serverless function or launch a VM or container in the Cloud. 
\end{itemize}

The Trigger life-cycle is as follows: An event is produced at some source. The event is consumed by the system, which \textbf{activates} the matching trigger. The event is processed by the Condition function. If the Condition results to be positive, then the event is processed by the Action function. When the Action is executed, we consider that the trigger has been \textbf{fired}. When a trigger has been fired, it can be disabled or maintained in the system, depending on if the trigger is configured as transient or persistent.

\textbf{Definition 3. Mapping workflows to triggers:} A workflow can be mapped to a set of Triggers ($\Delta$) which contains all state transitions ($\delta$ triggers) in the State Machine.  

We will show in next sections how different workflows (Amazon Step Functions) and Directed Acyclic Graphs (Apache Airflow)  can be transformed to a set of triggers ($\Delta$), which is the information needed by the Trigger service to orchestrate them.

For example, to transform a DAG into triggers, a trigger is added for every edge (workflow transition) of the graph. In a DAG, every node has its own unique ID, so the termination event from a task will contain as subject its ID to fire the trigger that handles its termination and invokes the next step in the workflow.

Thanks to the extensibility of the trigger architecture, any workflow abstraction that can be expressed as a Finite State Machine, can be translated into triggers and orchestrated by Triggerflow. For example, Triggerflow could orchestrate DAGs defined in other Domain Specific Languages (DSL) like DAX or Common Workflow Language (CWL), but a syntactic parser is needed for translation of those workflow DSL to triggers that can be interpreted and operated by Triggerflow. Also, basic triggers can be used to build more complex or specialized workflows that fit in a event-based and asynchronous scenario. For example, a set of custom triggers can be configured to pre-process or filter intermittent events that are originated from sensors. In Section \ref{sec:federated_learning} we describe a custom workflow to orchestrate a Federated Learning pipeline.

\textbf{Definition 4. Substitution principle:} A Workflow  must comply with an Action according to triggering (initialization) and finalization (Termination Event). A homogeneous treatment of Workflows and Actions permits nested workflow composition and iterations.

%Let us show how we can represent workflow operators using triggers. Let us represent trigger with two main calls addTrigger (event,Action) and addTrigger(Condition,Action). We will assume here that $A1^{T}$ is the Termination event of Action A1.

%To transform a DAG into triggers, a trigger is added for every edge (workflow transition) of the graph. In a DAG, every node has its own unique ID, so the termination event from a task will contain as subject its ID to fire the trigger that handles its termination and invokes the next step in the workflow. For example:

%\textbf{Sequence(A1,A2,A3,A4)}: Sequences can be created using simple addTrigger(Event,Action) invokations. In this case, the sequence can be transformed to: \newline addTrigger$(A1^{T},A2)$;addTrigger$(A2^{T},A3)$;addTrigger$(A3^{T},A4)$

%\textbf{Branching(if A1.result \textgreater 2 than 3 then A2 else A3)}: Branching require the condition engine. This can be converted to: \newline
%addTrigger$(A1,A1^{T}.result > 3,A2)$;addTrigger$(A1,A1^{T}.result > 3,A3);$

%\textbf{Fan-out(A1 $\rightarrow$  (A2,A3,A4))}: Again, fan-out do not require the condition engine: addTrigger$(A1^{T},[A2,A3,A4])$

%\textbf{Fan-in (A1,A2,A3) $\rightarrow$  A4)}: this now requires a conjunction conditional:
%addTrigger$('A1^{T} and A2^{T} and A3^{T}{'},A4)$

%These are simple examples of what we can achieve with a simple event algebra in conditions, but more sophisticated use cases could be presented by the extensible Condition engine.

\textbf{Definition 5.  Dynamic Trigger interception:} Any trigger can be intercepted dynamically and transparently to execute a desired action. Interception code is also performed with triggers. It must be possible to intercept triggers by condition identifier or by trigger identifier.  The condition identifier represents each existing condition in Triggerflow, for example a  map condition that aggregates all events in a parallel invocation. The trigger identifier represents the unique ID that each trigger receives on creation. %All trigger IDs and conditions are available in the Context.

%An example of intercepting a condition identifier could be a big data pipeline where we want to prewarm all functions executed in maps to improve system concurrency. For example : \textbf{addTrigger(map-condition, prewarm-action)}.  Condition id here is map-condition and refers to the instances of all triggers using this condition  type.
%An example of intercepting a trigger identifier could also be a Big data pipeline where we want to perform data prefetching or caching for a specific trigger in the workflow before the preparation phase. For example:\newline \textbf{addTrigger(preparationTriggerId,prefetch-action)} executes the prefetch action just before the preparationTrigger is activated.

%Dynamic trigger interception relies on computational reflection. 
We can introspect workflows, triggers, conditions, and actions using the Context. And we can intercept any trigger in the system in a transparent way using the Rich Trigger API. This opens the system to customize code and data in a very granular way. 
%It also enables the creation of  domain-specific algorithms performing smart provisioning policies to customize the size and shape of Cloud resources.

\subsection{Benefits and tradeoffs of event-based orchestration}

Event-based and reactive orchestration might not be the natural way to orchestrate a workflow. However, there are some benefits of using this approach that make event-based orchestration viable. First, the event bus is decoupled from the orchestration system, which is better suited for Cloud environments. This also simplifies the fault tolerance of the system if the event bus can resend uncommitted events. Also, we can leverage event bus services available in the Cloud like SQS on AWS that automatically scale on demand and and provide pay-per-use billing. By using events, we can reactively provision the orchestrator when events are produced, meaning that Triggerflow can autoscale to zero and only have allocated resources when state transitions take place. Finally, events are commonly used for service integration. Although Triggerflow is oriented mainly to FaaS orchestration, it can also orchestrate other services in the Cloud if they produce events, like containers that are executed in Container as a Service (like AWS Fargate) or a batch job running in a VM that has finished (like AWS Batch on self-managed EC2 instances).

The main disadvantage of using events and triggers for workflow orchestration is that Triggerflow only acts as a control plane. Task control and data flow are delegated to the application that is being orchestrated by Triggerflow. For example, when using serverless functions, the application has to rely on disaggregated storage services (like AWS S3) to pass data between tasks, because events are not meant to send large pieces of data. Also, debuggability is commonly poor in event-based systems. Triggerflow offers an event log and event replay as options to debug a workflow. However, many of these problems could be solved by running a workflow management engine (like Pegasus, Airflow, Nextflow, Argo...) on top of Triggerflow. These systems would manage the task and data plane while delegating the control plane to Triggerflow, thus benefiting from a reactive, scalable and resource-efficient orchestration in addition to the tools offered by the workflow management engine (GUIs, monitoring, logging...).

\subsection{Fault tolerance}
\label{sec:fault_tolerance}

In order to make Triggerflow tolerant to failures, we rely on the fault tolerance of the infrastructure where Triggerflow is deployed, the eventing service and the database. Regarding deployment fault tolerance, each system handles it differently, so it is explained in the next corresponding sections.

The event bus is required to guarantee at-least-once delivery. With at-least-once delivery, events can be duplicated and unordered. Triggerflow uses the CloudEvent standard, which includes a unique ID tag for every event. Repeated events with the same ID are discarded at the event consuming phase. Dealing with unordered messages depends on the kind of event composition that is taking place. In general, we can distinguish two types of event composition: aggregation and sequence. For aggregation, for example, a counter, the order of the messages does not alter the final result. For sequence, only events that activate the trigger that is at the head of the sequence are processed, other events are delayed until the triggers that they activate are enabled. For example, in the sequence $A \rightarrow B$, only the trigger A is enabled at first. If the event that activates B is consumed first, it is put into a Dead Letter Queue (DLQ), since B is disabled. Eventually, the event that activates A will be consumed, which activates and fires trigger A. Once trigger A is fired, trigger B is enabled and events on the DLQ will be processed again, this time activating correctly trigger B.

Each time a trigger is fired, a checkpoint of the current workflow state is persisted in storage: all contexts from triggers that have been activated are stored to the database and all events consumed until that moment are committed to the event broker. For example, if the system fails in mid of an aggregation event composition, the trigger will have been activated multiple times but not fired, so its state hasn't been checkpointed. At system restart, the event broker will send again uncommitted events, so the state will be eventually be restored as it was before the system failure. In this regard, trigger conditions are evaluated multiple times, while trigger actions are executed only once. So, the condition function is required to be \textit{idempotent}. Also, the database is required to be consistent and highly available.

\section{Prototype Implementation}
\label{sec:implementation}

Triggerflow has been implemented with Python $3.8$ for the application layer and Go $1.15$ for the system layer. Triggerflow can be deployed on Kubernetes along with two kind of autoscalers: one is Knative, which follows a push-based mechanism to pass the events from the event source to the appropriate worker, and another one with Kubernetes Event-driven Autoscaling (KEDA), where the worker follows a pull-based mechanism to retrieve the events directly from the event source. We created the prototypes on top of the IBM Cloud infrastructure, leveraging the services in its catalog to deploy the different components of our architecture. These components are the following:

\begin{itemize}
\item A Front-end RESTful API, where a user connects to interact with Triggerflow.
\item A Database, responsible for storing workflow information, such as triggers, context, etc.
\item A Controller, responsible for creating the workflow workers in Kubernetes.
\item The workflow workers (TF-Worker hereafter), responsible for processing the events by checking the triggers' conditions, and applying the actions.
\end{itemize} 

In our implementation, each workflow has its own TF-Worker. In other words, the scalability of the system is provided at workflow-level and not at TF-Worker level. In the validation (Sec. \ref{sec:validation}), we demonstrate how each TF-Worker provides enough event ingestion rate to process large amounts of events per second. %In this sense, we decided not to scale-up a TF-Worker of a given workflow more than once.

% The TF-Worker scales from 0 to 1, and from 1 to 0 depending on if there are events to process or not. 

%1) The FaaS services based on IBM Cloud Functions (IBM CF) and Knative. 2) A customized functions runtime, which generates termination events to the desired message broker. 3) A python client, which allows users to add and delete all the Triggerflow resources, such as event sources, triggers, and workspaces. 4) A front-end API based on IBM CF web actions \cite{ibmcf_webactions}. 5) A persistent database based on CouchDB. 6) The message brokers or event-sources based on Apache kafka. 7) The trigger service, which processes all the events of a given workspace and evaluates all the conditions of the triggers, eventually applying the actions, and 8) The controller, that manages (create or delete) the trigger service workers.

In our system, the events are logically grouped in what we call \emph{workflows}. The \emph{workflow} abstraction is useful, for example, to differentiate and isolate the events from multiple workflows, allowing to share a common context among the (related) events.

\subsection{Deployment on Knative}
\label{sec:deployment_Knative}

We mainly benefit from the Knative auto-scaler component in Knative Serving and the routing/filtering service in Knative Eventing.

Any serverless reactive architecture requires a managed multi-tenant component that is constantly running, monitoring event sources, and only launching actions in response to specific events. In this way, the tenant only pays for the execution of actions in response to events, and not for the constant monitoring of event services. For example, in OpenWhisk, when we create a trigger for a Function (like an Object Storage trigger), the system is in charge of monitoring the event source and only launching the function in response to events. 

In Knative Eventing, each tenant will have an Event Source that receives all events they are interested in (and have access to). We register a Knative Eventing Trigger for each workflow in the system. The filtering capabilities of Knative Eventing Triggers permit to route events of this workflow to the appropriate TF-Worker, which will activate the corresponding Triggerflow triggers.

Each event is tagged with a unique workflow identifier. We have created a customized functions runtime, which generates function termination events to the desired message stream that include the selected workflow identifier. If Triggerflow must receive events from services which do not include this workflow ID, a generic filtering service will match conditions to the incoming event (like ``all events of this object storage bucket belong to this workflow"), tag the event, and route it to the tenant's Event Source.

As each event contains a unique identifier per workflow, it is easy for Knative Eventing to route this event to the selected TF-Worker. The TF-Worker is then launched by Knative Serving to process the event, but it will also scale to zero if no more events are produced in a period. This ensures the serverless scale to zero and pay-as-you-go qualities for our Triggerflow service. %The TF-Worker accesses workflow state in the Context persistent store, which is also used for checkpointing and fault tolerance. 

Regarding \textbf{fault tolerance}, Knative Eventing guarantees ``at least once" message delivery, and automatic detection and restart of failed workers. %If a TF-Worker fails, the persistent Context will restore the state in a consistent manner after the failure. %The persistent Context is also used for stateful Conditions, like aggregation fork-join triggers that perform composite event detection and event counting.

%Developers declare in CRs description of their application components including autoscaling, retrying, and error handling (for example using "dead letter channels" to declare where undelivered events should go) and leave implementation details to the underlying event routing system.
%The major limitation of Knative Eventing in version 0.13 is performance - the project has not reached 1.0 and is not recommended for production use. As it currently lacks pull-based event monitoring mechanisms to instantiate containers in reaction to events, a Knative Eventing subscriber receives events in a Kubernetes pod through HTTP requests. This push-based mechanism adds overheads of translating events into HTTP and as events are not sent in batches the performance is affected compared to directly retrieving them from Kafka.
%substantially limits the number of events that a single pod can process to less than $1200$ e/second.

\subsection{Deployment on KEDA}

\begin{figure}[h]
	\includegraphics[width=0.45\textwidth]{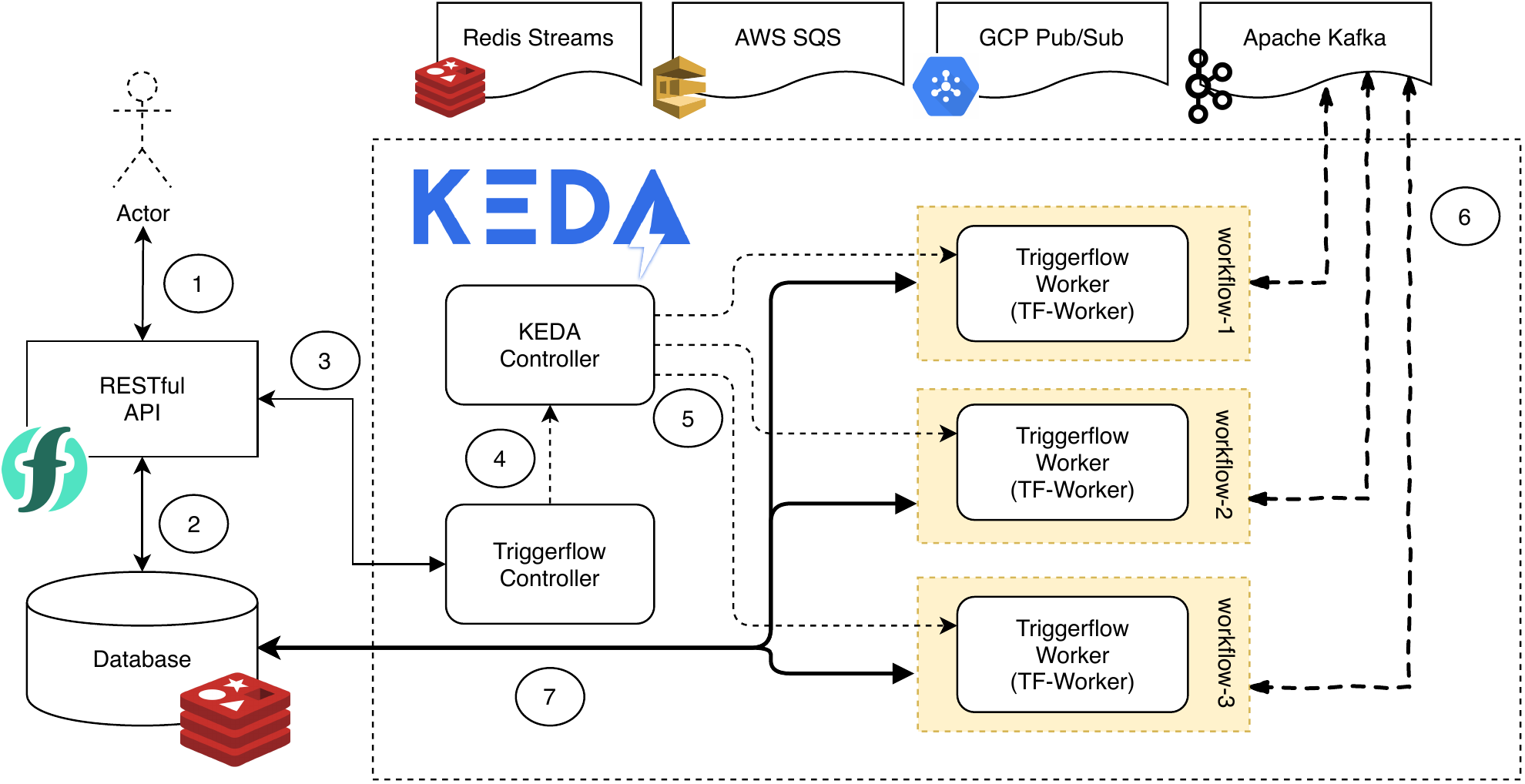}
	\centering
	\caption{Prototype deployment on KEDA}
	\label{fig:impl_keda}
	\vspace{-8pt}
\end{figure}

One of the hardest problems in event-driven applications is to deal with reliability and scalability. Event systems may be receiving events as soon as they are created (``pushed'') or they may process them when they are ready (``pull'' or ``poll'') and for both cases they need to deal with capacity limits and error handling. Knative is very well suited for push-based scaling as it can auto-scale based on incoming HTTP requests containing events. Kubernetes Event-driven Autoscaling (KEDA) is the best option now for event-based configurable pull-based scaling.

We have also implemented Triggerflow entirely on top of Kubernetes using the KEDA project \cite{keda}. KEDA offers pull-based configurable event queue monitoring and reactive scalable instantiation of Kubernetes containers. KEDA also offers configurable auto-scaling mechanisms to scale up or down to zero.

In this case, the Triggerflow Controller integrates KEDA for the monitoring of Event Sources and for launching the appropriate TF-Workers, and scaling them to zero when necessary. It is also possible to configure different parameters in KEDA like the queue polling interval, scale-out interval, and number of events scaling interval. Different types of workflows may require different configuration parameters.

The advantage here is that our TF-Workers connect directly to the event stream (Kafka, Redis Streams) using the native protocol of the platform. This permits to handle more events per second in a single pod. In contrast, with the Knative implementation, Knative Channels and Subscriptions are used. Knative Eventing consumes the event from the stream and then routes it via HTTP request to the corresponding TF-Worker Knative service. As we demonstrate in the validation, using KEDA allows us to handle intensive workloads from scientific workflows coordinating parallel jobs over thousands of serverless functions.

%To overcome the aforementioned performance problems of push-based eventing in Knative Eventing 0.13, we have also implemented Triggerflow entirely on top of  Kubernetes using the KEDA project \cite{keda}. KEDA offers pull-based configurable event queue monitoring and reactive scalable instantiation of Kubernetes containers. KEDA also offers configurable auto-scaling mechanisms to scale up or down to zero.

%In this case, the  Triggerflow Controller integrates KEDA for the monitoring of Event Sources and for launching the appropriate TF-Workers, and scaling them to zero when necessary. It is also possible to configure different parameters in KEDA like the queue pulling interval, passivation interval, and number of events scaling interval. Different types of workflows may require different configuration parameters.

%The advantage here is that, unlike in Knative Eventing, our TF-Workers connect directly to the Message Broker (Kafka, Redis Streams) using the native protocol of the broker. As events are sent in batches and do not have to be sent over HTTP the performance is much improved compared to the push-based approach described in previous section. As we will demonstrate in the validation, this allows us to handle intensive workloads from scientific workflows coordinating parallel jobs over thousands of serverless functions.

Figure \ref{fig:impl_keda} shows a high-level perspective of our implementation using KEDA. In this deployment, Triggerflow works as follows: through the client, an user must firstly create an empty workflow to the Triggerflow registry, and reference an event source that this workflow will use. Then, the user can start adding triggers to it (1). All the information is persisted in the database (for example, Redis) (2). Then, immediately after creating the \emph{workflow}, the front-end API communicates with the \emph{Triggerflow controller} (3), deployed as a single stateless pod container (service) in Kubernetes, to create the auto-scalable \emph{TF-Worker} in KEDA (4). From this moment, KEDA is responsible to scale up and down the TF-Workers (5). In KEDA, as stated above, the TF-Worker is responsible for communicating directly to the event source (6) to pull the incoming events. Finally, TF-Workers periodically interact with the database (7) to keep the local cache of available triggers updated, and to store the context (checkpointing) for fault-tolerance purposes.

%Regarding \textbf{fault tolerance}, we also guarantee ``at least once" message delivery and restarting of failed workers. In this case, the TF-Worker uses batching to commit groups of events in the Kafka Event Source once they have been correctly processed. If the TF-Worker fails, Kafka will just resend the non-committed events to the TF-Worker and thus ensuring message delivery.

Regarding \textbf{fault tolerance}, message delivery policies are now guaranteed by the messaging middleware. For example, Kafka guarantees that no messages are lost while $N - 1$ topic replicas are available. The Kubernetes scheduler will also restart failed workers. In this case, the TF-Worker uses batching to commit groups of events to Kafka once they have been correctly processed. If the TF-Worker fails, Kafka will just resend the non-committed events to the TF-Worker and thus ensuring message delivery.

In our Redis implementation, we use Redis both as event stream (Redis Streams), and as persistent store (for the Context and events). Again, if the TF-Worker fails, all events are in the event store, so it will continue with the non-processed events.

Currently, some experimental work \cite{keda-knative} is being done to incorporate KEDA autoscaler to Knative Event Sources components. Then, we would be able to deploy Triggerflow directly on top of one unified event router technology. It is also possible that some building blocks of Triggerflow could be moved to the Knative Eventing kernel. For example, the Knative Eventing community is now considering more advanced filtering mechanisms (complex event processing). In that case, our TF-Worker could delegate many tasks to the underlying event router.

\section{Use cases}
\label{sec:use_cases}

To demonstrate the flexibility that can be achieved using triggers with programmable conditions and actions, we have implemented three different workflow models that use Triggerflow as the underlying serverless and scalable workflow orchestrator.

\subsection{Directed Acyclic Graphs}

When a workflow is described as a Directed Acyclic Graph (DAG), the vertices of the graph represent the tasks of the workflow and the edges represent the dependencies between tasks. The fact that a DAG does not have cycles implies that there are no cyclic dependencies, which would be impossible to fulfill. 

The orchestration platforms that rely on DAGs for their workflow description, such as Apache Airflow, handle the dependencies between tasks with their \textit{downstream relatives} attribute, i.e. upon a completion of a task execution, these orchestrators look for what tasks have to be executed after the completed task.

However, from a trigger-based and reactive orchestration perspective, it is more compelling to know what tasks have to be executed before a certain one, i.e. what are the dependencies of every task, their \textit{upstream relatives}. With this information, we can register a trigger to activate a task's execution when all termination events from its upstream relatives are present.

If we assume that, upon a task completion, a termination event containing the ID of the completed task is produced, then we can orchestrate a DAG by adding a trigger for every vertex (task) with the following information:
\begin{itemize}
	\item As \textbf{activation events} of the trigger, we will register the termination events that are produced by tasks that execute before the current task, i.e. its dependencies.
	\item As \textbf{condition}, we count the number of events the trigger has to aggregate before executing the current task (for example, a join of a map execution or the number of branches to join).
	\item As \textbf{action}, we register the actual task to be executed, ideally an asynchronous task such as an invocation of a serverless function.
\end{itemize}

\begin{figure}[h]
	\includegraphics[width=0.45\textwidth]{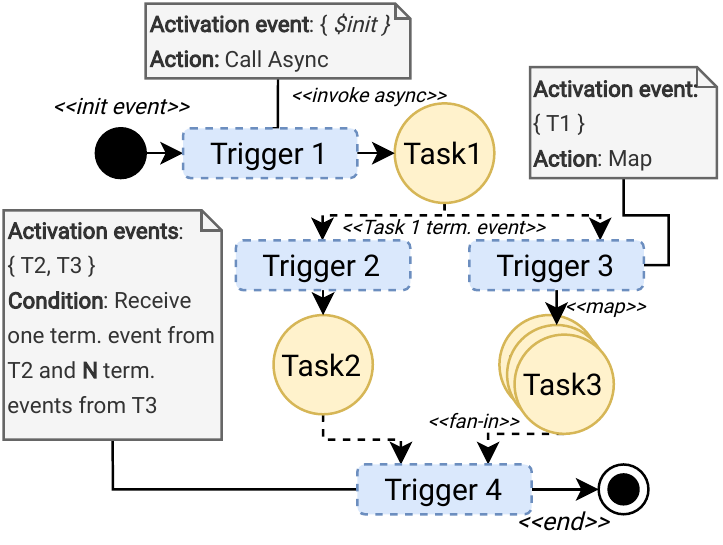}
	\centering
	\caption{Triggers that connect the tasks of an example DAG}
	\label{fig:dag2triggers}
	\vspace{-8pt}
\end{figure}

To handle a map-join trigger condition, before actually making the invocation requests, we use the \textit{introspect} context feature from the activated trigger action to dynamically modify the condition of the trigger that will aggregate the events, to set the specific number of expected functions to be joined. This is used in the case that the iterator which we map onto has a variable length depending on the workflow execution.

Furthermore, this approach gives us the opportunity to handle errors during a workflow runtime. Special triggers can be added that activate when a task fails, so that the trigger action can handle the task's error and halt the workflow execution until the error is solved. After error resolution (retry, skip or try-catch logic), the workflow's execution can be resumed by activating the corresponding trigger that would have been executed in the first place, as if there had not been an error.

%The implementation of the DAGs interface is inspired by Airflow's DAG definition, based on the premise of ``DAG definition as code'', which improves the readability, extensibility and reutilizability of workflows, aside from being much more user-friendly and comfortable to develop than other cluttery workflow definition objects based on metalanguages like JSON or YAML.

The DAGs interface implementation is inspired by Airflow's extensible DAG definition based on the  \textit{Operator} abstraction. According to Airflow's core ideas, an \textit{Operator} describes what is the actual work logic that is carried out by a task. Airflow offers a wide variety of operators to work with out of the box, but it can be extended through the implementation of \textit{plugins}. This approach is well suited to Triggerflow's architecture, thanks to its flexible programmatic trigger actions and conditions.

To illustrate this approach, Figure \ref{fig:dag2triggers} depicts how a simple DAG with \textit{call async}, \textit{maps}, and \textit{branches} is orchestrated using triggers.

\subsection{State Machines and Nested Workflows}

Amazon Step Functions bases its workflow description on a state machine defined by a declarative JSON object using the Amazon States Language DSL.

Similarly to Airflow's DAGs, a state machine definition in Amazon States Language (ASL) only takes into consideration what is the \textit{next} state to execute for each of them. However, from a trigger perspective, it is needed to figure out what states need to be executed before a given one. Then, we can add a trigger for every state transition that is activated by state termination events and handles the state machine flow logic. 

Nevertheless, a distinctive feature that ASL provides is that a state can be a sub-state machine. For instance, the primitives \textit{map} and \textit{parallel}, map and branch to an entire state machine, rather than a single task like in the DAG interface. To manage this feature, we need a special event that is produced when a state machine ends. For map and branch joins, we will then join those sub-state machines instead of single states. To do so, we identify each sub-state machine with a unique tag in the scope of the execution. By doing so, we also comply with the substitution principle of the serverless trilemma.

To produce state machine termination events, we need to activate triggers from within a trigger action/condition function, as state machine joining is detected in there. To do so, the worker's event sink internal buffer was made accessible through the context object so that a trigger action/condition function can internally produce events that activate the necessary subsequent triggers.

In an Amazon Step Functions execution, the states can transfer their output to the input of the following state. To reproduce this functionality, we transfer data by passing it through the termination events. This way, the output of a state can be parsed from the consumed event in the trigger action and used as input for the following state.

If we consider a state machine to be itself a state, we can seamlessly compose ASL definitions in other state machines with its triggers and connections. Amazon Step Functions, however, is more limited in terms of task extensibility since we are given a closed set of state types. We will explain here how these are processed with triggers:

\begin{itemize}
    \item \textbf{Task} and \textbf{Pass} states: These state types carry out the actual workflow computational logic, the rest of the state types only manage the state machine flux. The Task state relies on the asynchronous Lambda invoked to signal the next trigger upon its termination, whereas the Pass state signals itself its termination event.
    \item \textbf{Choice} state: The choice state type defines a set of possible outcomes that execute depending on some basic boolean logic that can compare numbers, timestamps, and strings. The trigger approach for this state is simple: for all possible outcomes apply the condition defined in the Choice state to the condition field of the trigger that handles its state execution.
    \item \textbf{Parallel} state: This state type defines a set of sub-state machines that run in parallel. In this case, we will iterate each sub-state machine and collect their IDs. Finally, we add a trigger that is activated whenever any of those sub-state machines ends, but it is only executed when it has been signaled by every sub-state machine.
    \item \textbf{Map} state: Similarly to the Parallel state type, this state defines a single sub-state machine that executes for every element in an iterable data structure input in parallel. Before executing the sub-state machines, we first add a trigger that, during its action execution, checks the length of the iterable object (which is the number of parallel state machines, unknown until execution), and registers it to the trigger context that handles the sub-state machines termination stating how many of them it should wait for.
    \item \textbf{Wait} state: The Wait state type waits for a certain amount of seconds, or until a timestamp is reached before continuing. It can be implemented by registering the activation event production that activates the trigger to an external time-based scheduler.
    \item \textbf{Fail} and \textbf{Succeed} states: The Fail and Succeed states stop the execution of the state machine and determine if it executed successfully or failed. It can be implemented assigning special actions to their triggers that end the execution of the workflow.
\end{itemize}

\begin{figure}[h]
	\includegraphics[width=0.45\textwidth]{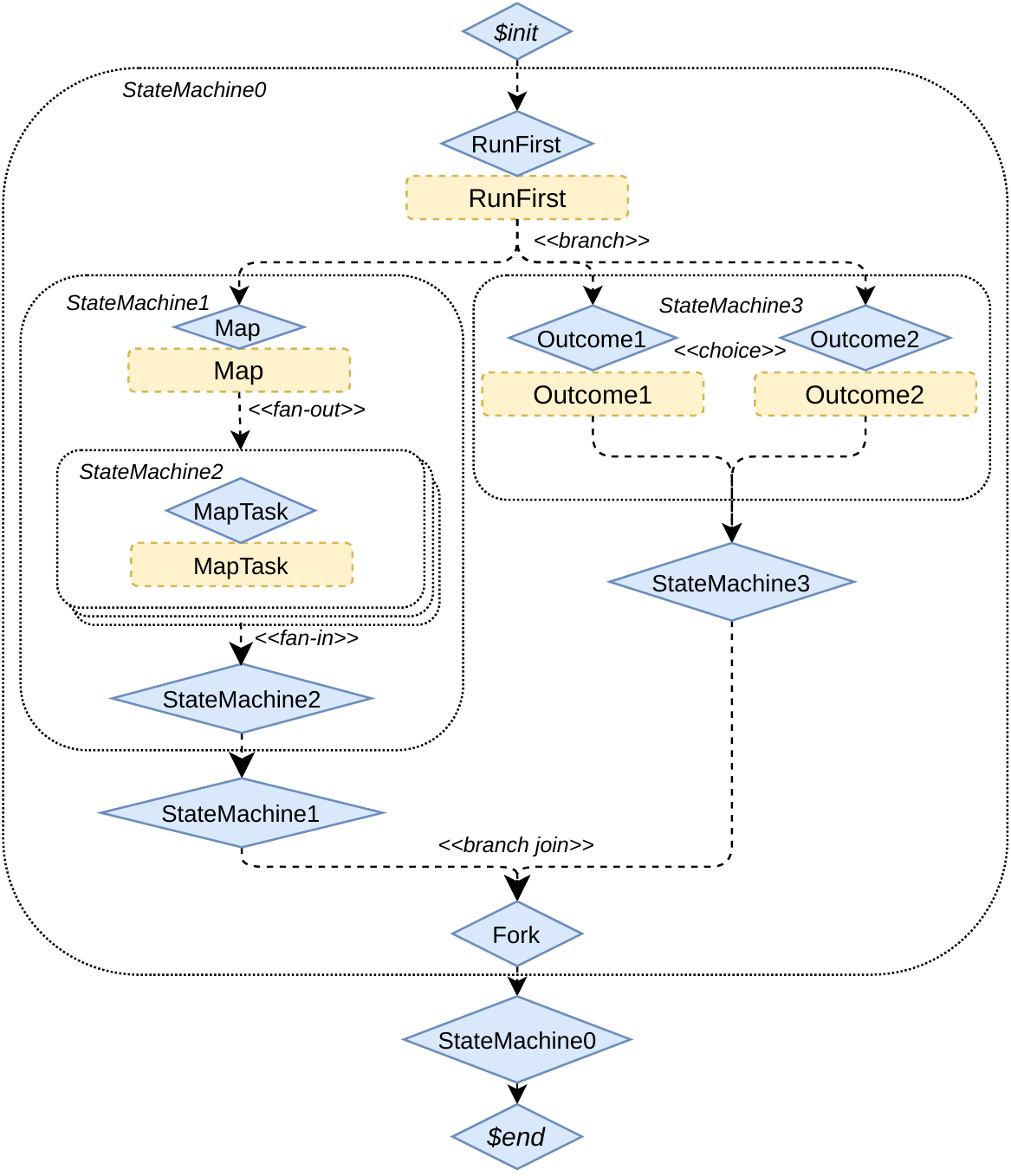}
	\centering
	\caption{Triggers representation of an ASF state machine}
	\label{fig:asf}
	\vspace{-8pt}
\end{figure}

Figure \ref{fig:asf} depicts how an ASF state machine is orchestrated by triggers.

\subsection{Workflow as Code and Event Sourcing}
\label{sec:workflow_as_code}

The trigger service is also useful to reactively invoke an external scheduler because of  state changes caused by some condition. For example, Workflow as Code systems like Lithops or Azure Durable Functions represent state transitions as asynchronous function calls (async/await) inside code written in Python or C\#.  Asynchronous invocations and futures in Lithops or  async/await calls in Azure Durable Functions simplify code so developers can write synchronous-like code that suspends and continues when events arrive. 

The model supported by Azure Durable Functions is reactive and event-based, and it relies on event sourcing to restart the function to its current state.  We can use dynamic triggers to support external schedulers like Durable Functions that suspend their execution until the next event arrives.  For example, let's look at this Lithops code:

\medskip
\begin{lstlisting}[language=Python]
import lithops

def my_function(x):
    return x + 3

lith = lithops.FunctionExecutor()
future = lith.call_async(my_function, 2)
result = future.result()  # result = 5
futures = lith.map(my_function, range(result))
print(lithops.get_result(futures))  # prints "[3, 4, 5, 6, 7]"
\end{lstlisting}

In this code, the functions \emph{call\_async} and \emph{map} are used to invoke one or many functions.
Lithops code like this is executed normally in the client in a notebook, which is usually adequate for short running workflows. But what if we want to execute a long-running workflow with Lithops in a reactive way? The solution is to run this Lithops code in Triggerflow reacting to events.  Here,  prior to perform any invocation, Lithops can register the appropriate triggers, for example:

\medskip
\textbf{call\_async(my\_function, 3)}: Inside this code we will dynamically register a function termination trigger.

\textbf{map(my\_function, range(res))}: Inside this code we will dynamically register an aggregate trigger for all functions in the map.

\medskip

After trigger registration for each function, the function can be invoked and the orchestrator function could decide to suspend itself. It will be later activated when the trigger fires.

To ensure that this Lithops code can be restarted and continue from the last point, we use \emph{event sourcing}. When the orchestrator code is launched,  an event sourcing action will re-run the code acquiring the results of functions from termination events. It will then be able to continue from the last point.

In our system prototype, the event sourcing is implemented in two different ways: native and external scheduler.

In the \emph{native scheduler}, the orchestration code is executed inside a Triggerflow Action. Our Triggerflow system enables then to upload  the entire orchestration code as an action that interacts with triggers in the system. When Triggerflow detects events that match a trigger, it awakens the native action. This code then relies on event sourcing to catch up with the correct state before continuing the execution.  In the native scheduler, the events can be retrieved efficiently from the context and thus accelerate the replay process. If no events are received in a period, the action will be scaled to zero. This guarantees reactive execution of event sourced code.

\begin{figure}[h]
	\includegraphics[width=0.45\textwidth]{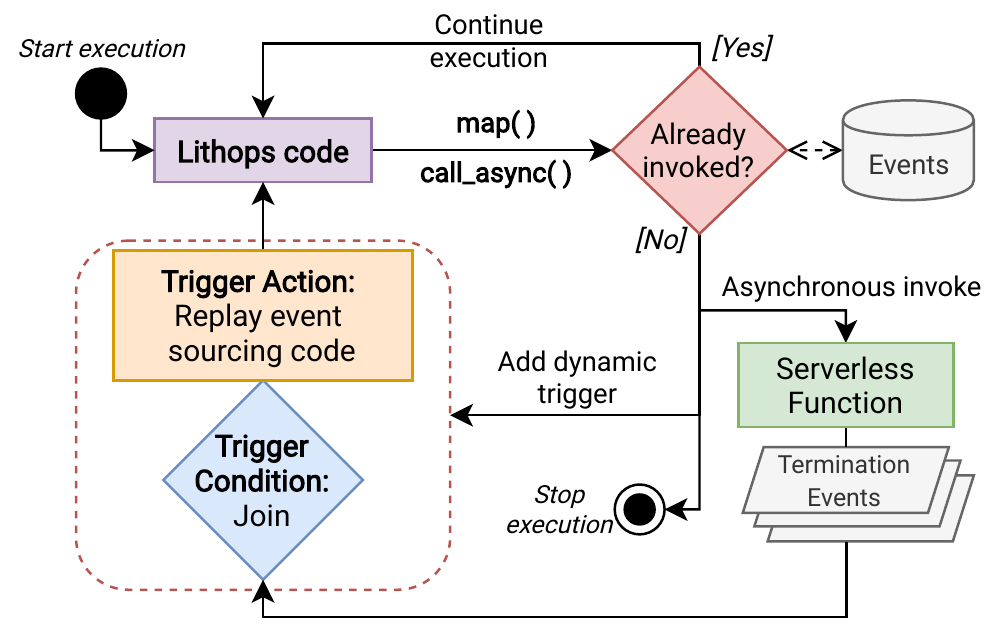}
	\centering
	\caption{Life cycle of an event sourcing-enabled workflow as code with Lithops as external scheduler.}
	\label{fig:event-source-Lithops}
	\vspace{-8pt}
\end{figure}

In the \emph{external scheduler}, we use Lithops Serverless Framework \cite{lithops}, where the orchestration code is run in an external system, like a Cloud Function. Then, thanks to our Triggerflow service, the function can stop its execution each time it invokes for example a \emph{map()}, recovering their state (event sourcing) when it is awaken by our TF-Worker once all \emph{map()} function activations finished their execution. Moreover, to use our event sourcing version of Lithops, it is not required any change in the user's code. This means that the code is completely portable between the local-machine and the Cloud, so users can decide where to run their Lithops workflows without requiring any modification. The life cycle of a workflow using an external scheduler can be seen in Figure \ref{fig:event-source-Lithops}.

\subsection{Specialized Workflows: Federated Learning Orchestrator}
\label{sec:federated_learning}

We have leveraged the flexibility of Triggerflow to implement a Federated Learning orchestrator using triggers. Federated learning consists of training a machine learning model in a distributed and iterative way, where each server or client trains the model using a local and private portion of the whole dataset. A central server acts as an orchestrator for the whole process: it is responsible for selecting the candidate clients for each round, transmit the initial model to each of them, and then wait for the clients to send back their trained models. It then aggregates the results generating a unique model. The central server can then decide whether the model is accurate enough and stop the process or to start another round to increase the model accuracy. In a federated learning scenario, clients are heterogeneous and may be unreliable, they don't know each other and they can't share data since that data could be sensitive or confidential (for example, clinical patient data from a hospital). Those clients participate in the training process in an unpredictable way, meaning that the total number of clients might fluctuate greatly during the process, as some of them can unexpectedly fail or leave the client pool at any given moment. A common methodology for the controller server is using a centralized architecture, but this approach does not scale.  Distributed architectures do scale, but at the expense of complicating fault tolerance. For example, in \cite{bonawitz2019towards}, the authors propose a distributed approach based on actors. However, they state that if an aggregator actor fails, the clients that are connected to it are lost, which leads to data loss. Either way, in both approaches the controller service is running during the whole training process, which might take several hours.

We can leverage Triggerflow's flexibility to build a custom workflow made of triggers that act as a loosely coupled fault-tolerant and serverless-like controller service for the Federated Learning process. The workflow is designed as a cyclic process using two triggers: the \textsl{aggregator} trigger, which controls training rounds and updates the model with the partially aggregated results, and the \textit{round} trigger, that decides when to restart the cycle and train another model. 

\begin{figure}[h]
	\vspace{-12pt}
	\includegraphics[width=0.45\textwidth]{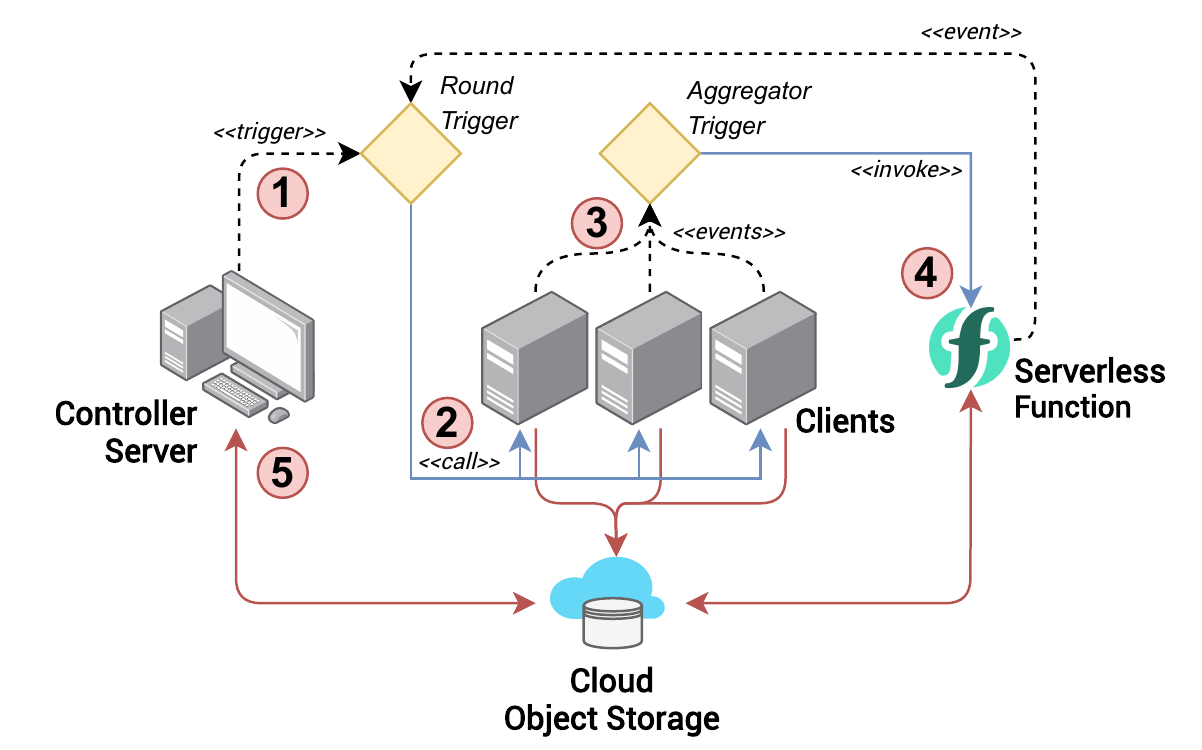}
	\centering
	\caption{Federated learning workflow orchestrator diagram.}
	\label{fig:federated}
	\vspace{-8pt}
\end{figure}

A diagram of the workflow is represented in Figure \ref{fig:federated}. First, the controller sets up the triggers with the corresponding model information to train in this round. It then triggers the \textit{round trigger} in order to start the first training round $(1)$. The \textit{round trigger} calls all the available clients to start training the model $(2)$. The clients then proceed to locally train the model and, upon end, they save the trained model weights to cloud object storage and send an event to the \textit{aggregator} trigger containing the object result key $(3)$.

The \textit{aggregator trigger} operates with a custom condition. Depending on the round and the number of clients, it waits for all clients or just a subset of them to send their termination event. This is used when some of the clients take a longer time to train the model, and the \textit{aggregator trigger} decides not to wait for them since it would slow the whole training process. We can also intercept the trigger with a timeout event produced by a cron job. This is useful when some or all clients leave the client pool, so the \textit{aggregator trigger} won't be waiting for them indefinitely. When the condition has aggregated all result keys from the selected clients, the action is fired, which invokes a serverless function that retrieves the model weights from object storage and performs a model update by aggregating the results of that round $(4)$.

After aggregating the trained deltas of all clients, the function stores the result on the cloud and deletes all the intermediate data stored in it. At last, it generates a completion event that is sent to the \textit{round} trigger.

The \textit{round} trigger is activated when the aggregator serverless function has aggregated all client models. It can then decide if the model is accurate enough or if another round has to take place to improve it. In that case, it would call the available clients with the updated model and the cycle would start again. If it decides that the training has finished, it can notify the controller server by, for example, sending a request to a specific endpoint, containing the final model $(5)$.

In contrast to a centralized architecture (like \cite{hard2019federated}), Triggerflow enables time and space decoupling and high scalability by design, as well as fault tolerance with event sourcing. Also, note that during the learning phase, the controller server can be deprovisioned to save compute resources, as all the orchestration process is offloaded to Triggerflow, which also auto-scales based on the events that are produced at the partial weights aggregation phase.

\section{Validation}
\label{sec:validation}

Our experimental testbed consists of $5$ client machines with $4$ CPUs and $16$ GB RAM. On the server side, we deploy Triggeflow on a Kubernetes installation (v1.17.3) in a rack of $5$ Dell PowerEdge R$430$ (2 CPUs Intel(R) Xeon(R) CPU E5-2620 v4 @ 2.10GHz - 8 Cores/CPU - 32 Logical processors) machines with $16$GB RAM. All of these machines, including the clients, are connected via 10GbE network, and run Ubuntu Server $19.04$. For the experiments we use Kafka 2.4.0 (Scala 2.13), RabbitMQ 3.8.9, Redis 5.0.7, KEDA 1.3.0 and Knative 0.12.0.

\subsection{Load test}

The load test objective is to demonstrate that our system can support high-volume event processing workloads in an efficient way. This is mandatory if we want to support the execution of high performance scientific workflows. 

For the first experiment, we want to measure how many events per second can be processed by a worker that consumes events from a message broker like Kafka or Redis Streams and the overhead produced in the trigger processing pipeline (stateful condition and action functions). Table \ref{tab:load_test} shows the time and throughput to process $200$K events in a container using different CPU resources ($0.25$, $0.5$, $1$ and $2$). \emph{Noop} means that the worker is not doing any operation on the event. \emph{Join} refers to $100$ triggers with aggregation filters that join $2000$ events each, resembling a multiple parallel map fork-join processing scenario. As we can see, the performance numbers tell that the system can process thousands of events per second with low overhead. Also, the system leverages multiple CPUs to increase the processing capacity.

\begin{figure}[h]
    \vspace{-12pt}
	\includegraphics[width=0.45\textwidth]{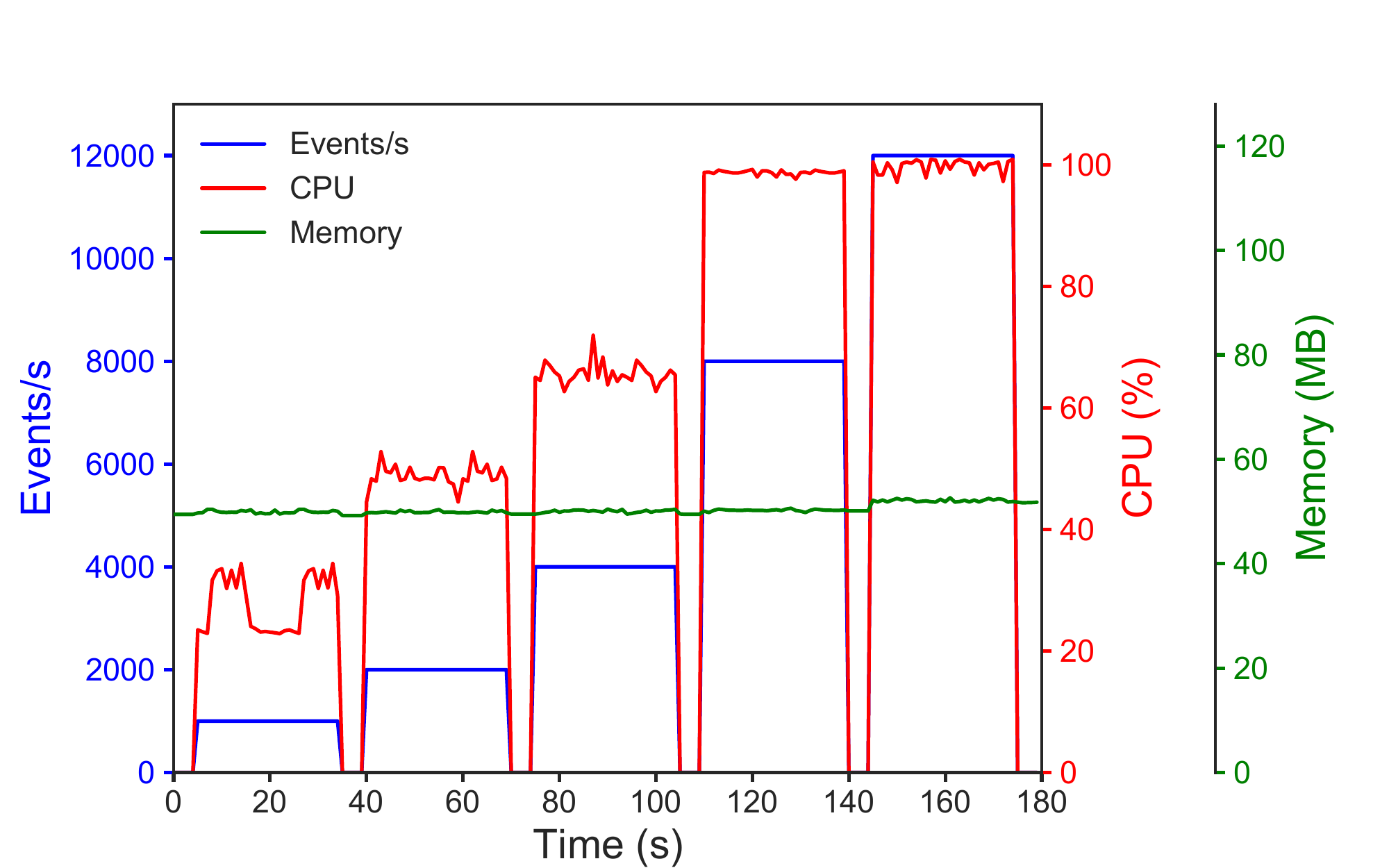}
	\centering
	\caption{Resource utilization depending on incoming number of events/second (1 Core w/ Redis)}
	\label{fig:load_test}
	\vspace{-8pt}
\end{figure}

The second experiment consists of measuring the actual resource usage (CPU and mem) of $1$ Core worker using Redis by injecting different numbers of events per second (form $1$K e/s to $12$K e/s). Figure \ref{fig:load_test} shows that, with a constant memory footprint, the CPU resource can cope with increasing number of events per second.

\begin{table}[t]
	\centering
	\begin{adjustbox}{width=0.5\textwidth}
	\begin{tabular}{@{} *7l @{}}
	\toprule
	\textbf{\makecell{Time\\elapsed (s)}} & \multicolumn{3}{c}{Noop}    & \multicolumn{3}{c}{Triggerflow}                 \\
	\midrule
	 CPU          & \emph{Redis} & \emph{Kafka} & \emph{RabbitMQ} & \emph{Redis} & \emph{Kafka} & \emph{RabbitMQ} \\ 
	 $2$          & $11.94$ & $2.64$  & $10.18$  & $12.15$ & $2.77$  & $10.38$  \\
	 $1$          & $11.97$ & $3.26$  & $10.08$  & $12.22$ & $4.10$  & $11.36$  \\
	 $0.5$        & $12.06$ & $5.02$  & $15.45$  & $12.88$ & $9.69$  & $24.66$  \\
	 $0.25$       & $12.85$ & $12.22$ & $39.46$  & $14.67$ & $20.79$ & $52.94$  \\
	\midrule
	\textbf{\makecell{Throughput\\(events/s)}} & \multicolumn{3}{c}{Noop}    & \multicolumn{3}{c}{Triggerflow}                 \\
	\midrule
	 CPU          & \emph{Redis} & \emph{Kafka} & \emph{RabbitMQ} & \emph{Redis} & \emph{Kafka} & \emph{RabbitMQ} \\ 
	 $2$          & $16743.87$ & $75665.85$  & $19639.03$  & $16460.87$ & $72188.87$ & $19261.42$  \\
	 $1$          & $16707.47$ & $61298.16$  & $19825.33$  & $16364.08$ & $48718.57$ & $17598.95$  \\
	 $0.5$        & $16584.17$ & $39818.74$  & $12942.01$  & $15525.39$ & $20632.44$ & $8110.25$  \\
	 $0.25$       & $15567.81$ & $16365.36$  & $5067.72$   & $13625.84$ & $9618.46$  & $3777.61$  \\
	\bottomrule
	 \hline
	\end{tabular}
\end{adjustbox}
\caption{Maximum number of processed events/second using Redis Streams, Kafka and RabbitMQ}
\label{tab:load_test}
\end{table}

\subsection{Auto-scaling}
In this case, the objective is to demonstrate that TF-Workers can scale up and down based on the current active workflows.
We demonstrate here that our Triggerflow implementation on top of Kubernetes and KEDA can auto-scale on demand based on the number of events received in different workflows.

For this experiment, we use the entire testbed described above, and set the TF-Worker to use $0.5$ CPUs and $256$ MB of RAM. The test consists of $100$ synthetic workflows that send events during some arbitrary seconds, pause the workflow for a while (simulating a long-running action), then resume sending events, and finally stop the workflow. The test works as follows: It first starts $50$ workflows at a constant rate of $2$ workflows per second), after $100$ seconds it starts another $50$ workflows at a rate of $3$ workflows per second, and finally, after $70$ seconds, it starts $15$ more workflows at a rate of also $3$ workflows per second.

\begin{figure}[h]
	\vspace{-12pt}
	\includegraphics[width=0.45\textwidth]{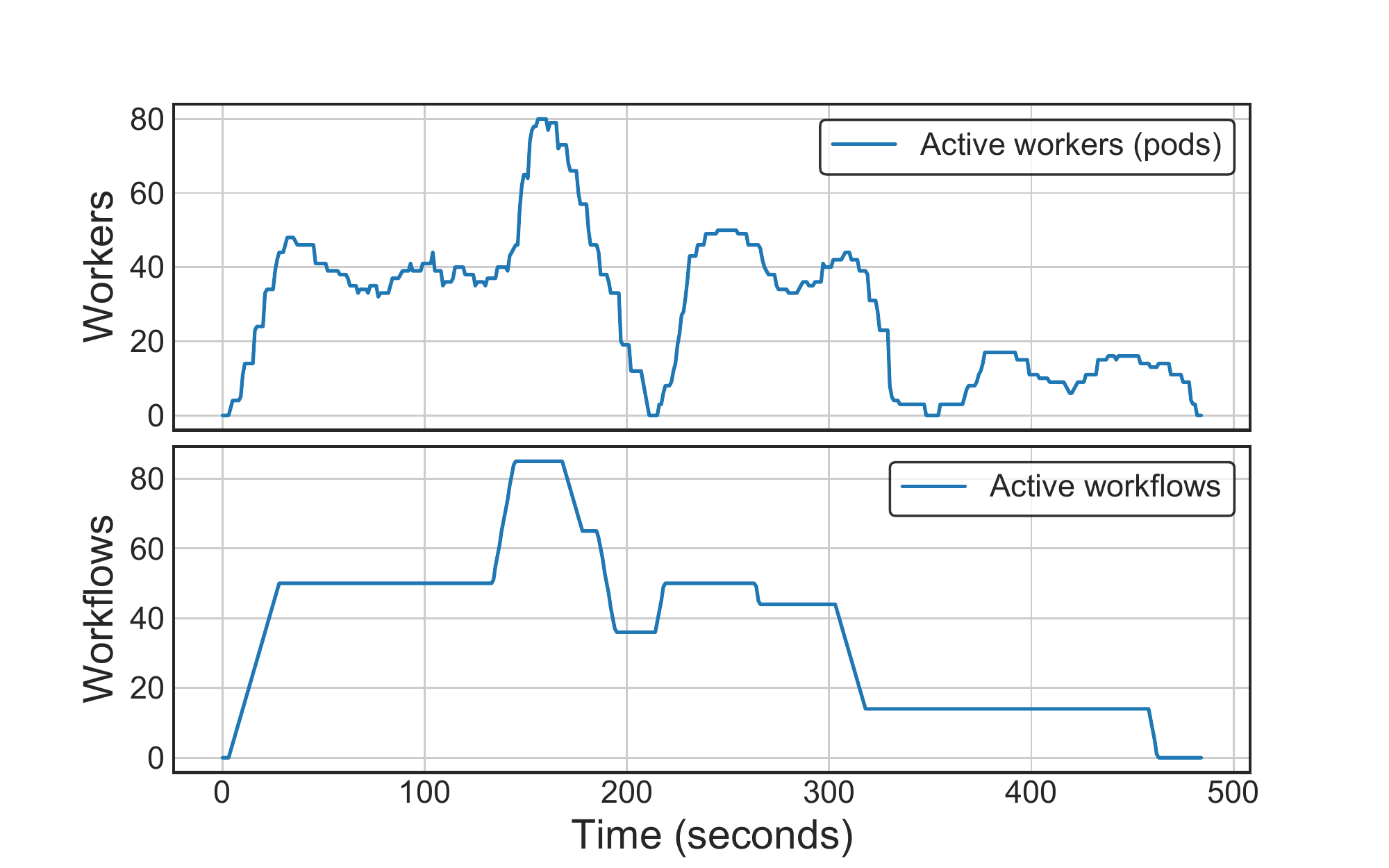}
	\centering
	\caption{TF-Worker auto-scaling test using KEDA}
	\label{fig:autoscaling}
	\vspace{-8pt}
\end{figure}

The results are depicted in Figure \ref{fig:autoscaling}. It shows how the TF-Workers scale up when the workflows start to send events, and scale down, even to zero (second 210 and 250), when the active workflows do not produce any event due to a long-running action. We can see how Triggerflow leverages the KEDA auto-scaler to activate or halt workflows. Triggerflow is automatically providing fault tolerance, event persistence, and context and state recovery each time a workflow is resumed.

\subsection{Completion time and overhead}

The validation in this section demonstrates that Triggerflow shows comparable overhead to public Cloud orchestration systems.
We must be fair here: we are comparing an implementation of Triggerflow over dedicated and idle resources in our rack against public multi-tenant cloud services that may be used by thousands of users. The objective is not to claim that our system is better than them, but only to demonstrate that we can reach comparable overhead and performance.
Furthermore, most cloud orchestration systems are not designed for highly concurrent and parallel jobs, which can limit their performance in those scenarios.

We evaluate the run-time overhead of Amazon's, IBM's, and Microsoft's orchestration services. We consider as \emph{overhead} all the time spent outside the functions being composed, which~is easy to measure
in all platforms. For a sequential composition $g$ of $n$ functions  $g = f_1  \circ f_2 \circ \dots \circ  f_n$, it is just:
 
\[
\mbox{overhead }({g})=  \mbox{exec\_time}( g) - \sum^n _{i = 1} \mbox{exec\_time}( f_i).
\]

It is important to note that our overhead definition includes the delays between function invocations, and the execution time of the orchestration function (for IBM Composer and ADF) or the delays between state transitions (for ASF). In the case of Triggerflow, the overhead depends on all the services in the architecture---i.e., latency to access Kafka or Redis, latency to invoke functions in IBM CF, etc.

For all the tests, we use a single TF-Worker with $0.5$ CPU Cores and $64$MB of RAM, and we list only the results when functions are in warm state. This implies that we do not consider the cold start of spawning the function containers and VMs. Our focus is on measuring the overhead of running function compositions. All the tests are repeated $10$ times. The results displayed are the median of those 10 samples and the standard deviation for the error intervals. Measurements are done during March of 2020. For IBM Cloud Functions (IBM CF) and AWS Lambda executions, we use the Python 3.8 runtime. The exception is Azure, which does not currently support Python for ADF, but C\#. The orchestration functions are implemented in the default language available in each platform: Node.js for IBM Composer, and C\# for ADF. ASF orchestration is specified in Amazon States Language (JSON-based format) using the console editor.
% IBM Sequences were statically defined by a command-line argument at deployment time.

For the \emph{sequential workflows}, we quantify the overhead for sequential compositions of length \emph{n} in \{5, 10, 20, 40, 80\}. For simplicity, all the functions in the sequence are the same: a function that sleeps for 3s, and then returns. For the \emph{parallel workflows}, we define a workflow with a single parallel stage composed of $n$ parallel instances of the same task, with $n$ ranging from $5$ to $320$, and doubling each time. This task has a fixed duration of $20$ seconds. Consequently, any execution of the experiment should ideally last $20$ seconds, irrespective of $n$ or the environment. To put it in another way, in an ideal system with no overhead, the execution time of the $n$ concurrent tasks should match that of a single task. Therefore, we compute the overhead of the orchestration system by subtracting the fixed time of a single task, namely $20$ seconds, from the total execution time.

\subsubsection{DAGs and State Machines}
For the DAG and State Machine use cases, we evaluated our DAG engine interface against IBM Composer, AWS Step Functions, AWS Step Functions Express, and Azure Durable Functions. It is important to state that these results are exactly the same we would get for the State Machine implementation over Triggerflow. Sequences and parallel jobs in state machines and DAGs use the same triggers.

\begin{figure}[h]
	\includegraphics[width=0.45\textwidth]{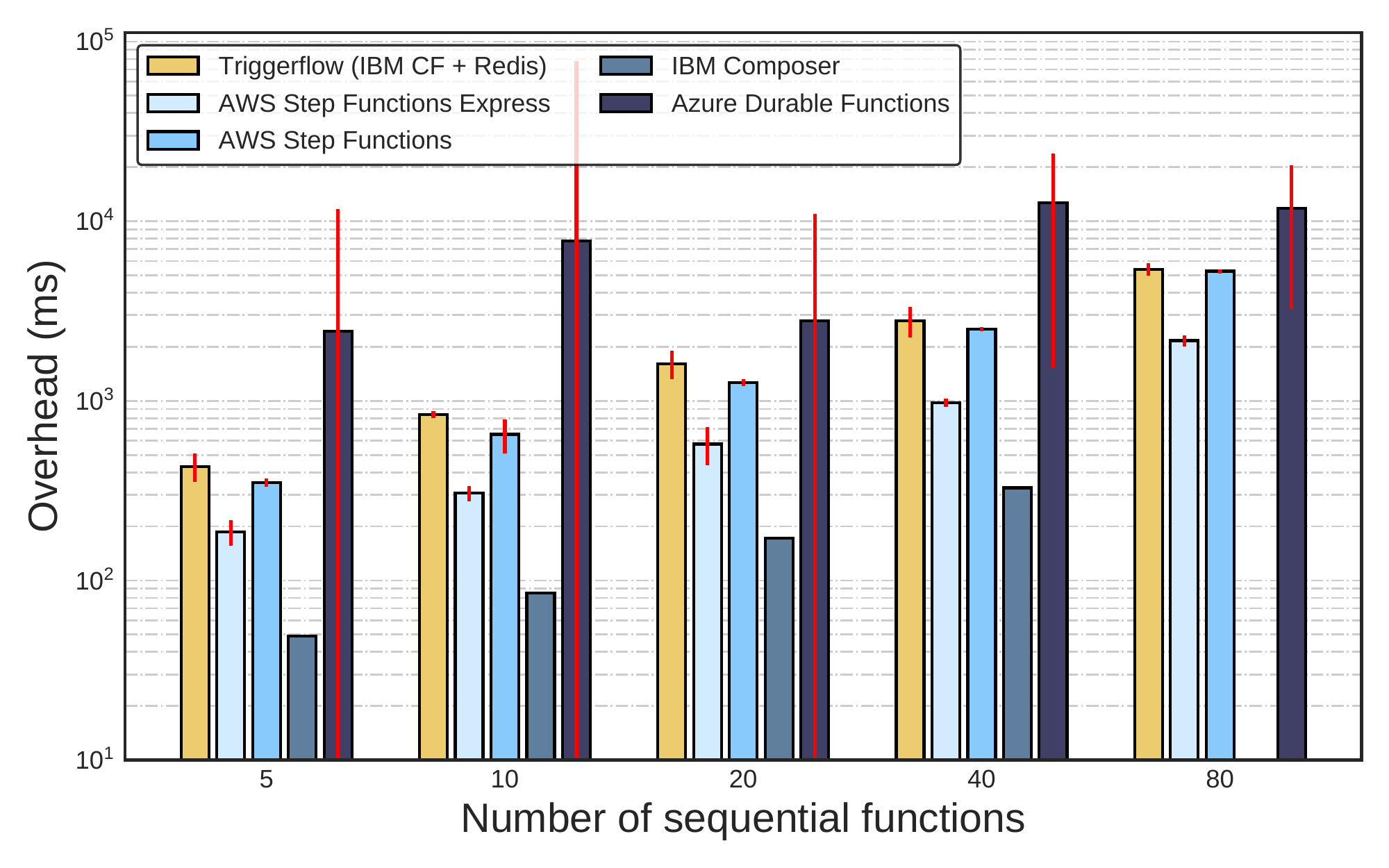}
	\centering
	\caption{DAG overhead comparison for sequences}
	\label{fig:overhead_sequence}
	\vspace{-8pt}
\end{figure}

\noindent{\textbf{Sequential workflows.}} The resultant overhead is represented in Figure \ref{fig:overhead_sequence}. In general, Triggerflow's overhead is comparable to Amazon Step Functions'. In this case, almost all overhead comes from the IBM Cloud Functions invocation latency using its public API, which is about $0.13$s. When multiplied by 80 functions, it adds up to approximately 10 seconds of overhead. Amazon Step Functions may be using internal trigger protocols rather than the public API, which should lower invocation latency. However, it is probably running in shared resources, which could explain the simmilarity in overhead. In addition, it seems that using Express Workflows does not provide a considerable speed improvement compared to regular ASF when using sequential workloads, so they are probably not worth the extra cost for this kind of job. IBM Composer is the fastest in sequences, but with the drawback of its limitation of only 50 transitions per composition. Finally, Azure Durable Functions present competent overheads for long sequences, although being quite unstable for short sequences. This is probably because ADF is designed and optimized for long-running sequential workloads.

\begin{figure}[h]
	\includegraphics[width=0.45\textwidth]{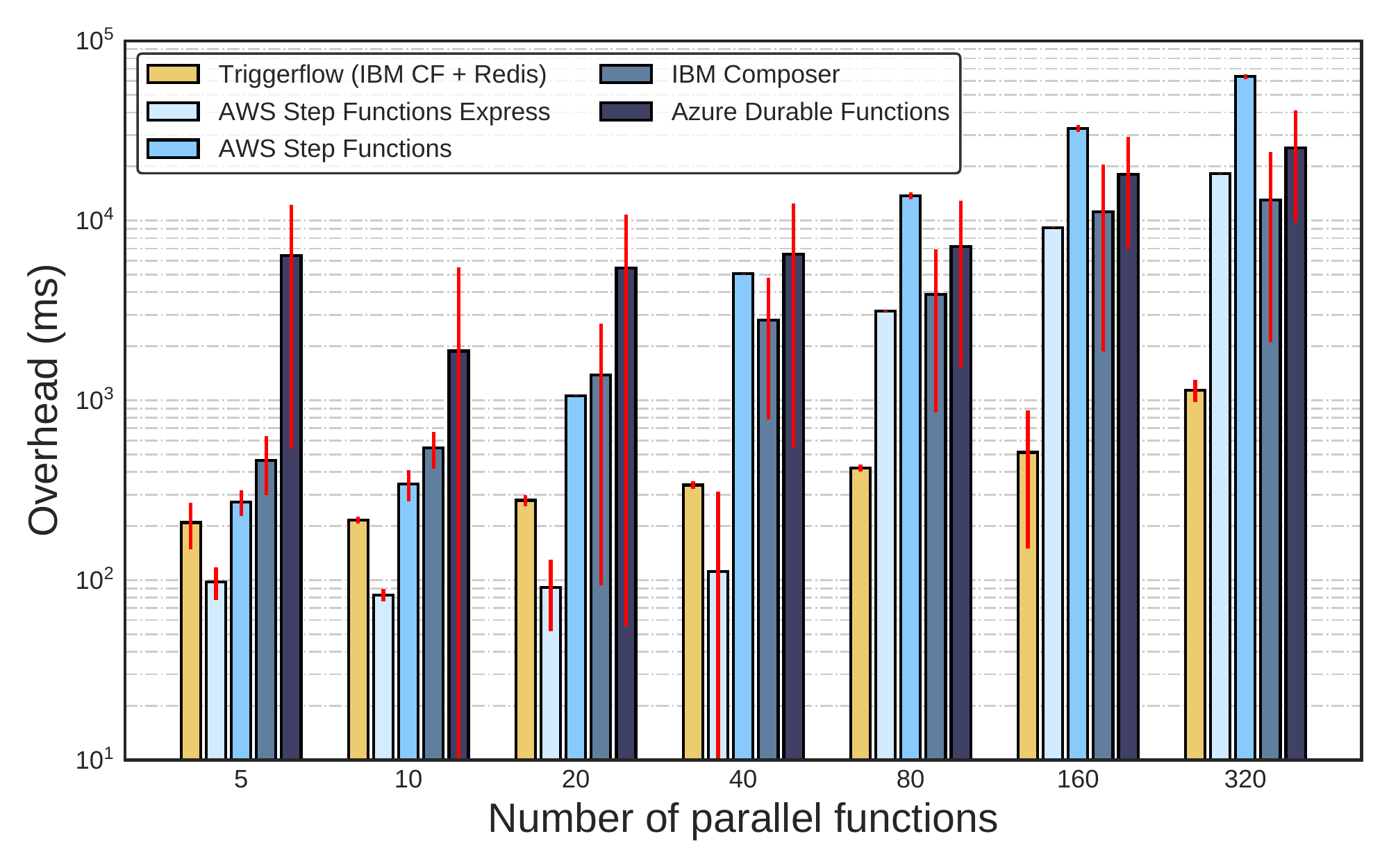}
	\centering
	\caption{DAG overhead comparison for parallel workflows}
	\label{fig:overhead_map}
	\vspace{-8pt}
\end{figure}

\medskip
\noindent{\textbf{Parallel workflows.}} For small-sized compositions (5 to 10), we can see in Figure \ref{fig:overhead_map} that Triggerflow and AWS Step Functions yield similar overhead, both being outperformed by Express Workflows nonetheless. Express Workflows has a wider range of error though, while regular Step Functions, Triggerflow and IBM Composer are more stable. Express Workflows perform similarly regardless of the number of parallel functions until it reaches about 80, when its performance drops drastically and the overhead skyrockets for no apparent reason. From 80 functions and up, Express Workflows and IBM Composer have similar overheads.

From 80 parallel functions and up, we also see that Triggerflow has the lowest overhead, proving that event-driven function composition is indeed well suited for large parallel map function joining.

Azure Durable Functions yield the worst results when used for small-sized function joining and is considerably unstable. However, it turns to be equivalent to the other orchestration systems when joining a higher number of concurrent functions.

\subsubsection{Workflow as Code and Event Sourcing}
The objective here is to evaluate Workflow as Code and event sourcing overheads in Triggerflow compared to Azure Durable Functions. We compare both sequential and parallel constructs.

For the event sourcing use case, we evaluate both the \emph{external scheduler} (Lithops) and the \emph{native scheduler} (Triggerflow action).
One the one hand, we measure and compare the performance of our modified version of Lithops for Triggerflow with the original version of Lithops (\emph{external scheduler}). In this case we evaluate 4 different scenarios: 1) The original Lithops, which makes use of IBM Cloud Object Storage (COS) to store the events and results. 2) The modified version of Lithops for Triggerflow that stores the results in COS (original Lithops behavior), but sends the termination events trough a Redis Stream. 3) The Triggerflow Lithops that sends the events and results trough a Kafka Topic. And 4) the Triggerflow Lithops that sends the events and results trough a Redis Stream.

On the other hand, we evaluate the native Triggerflow event sourcing scheduler, where the orchestration code is executed as part of the trigger action. In this case we compare the results against the Azure Durable Functions (ADF) service, which is the only FaaS workflow orchestration service that employs an event sourcing technique to execute the workflows.

\begin{figure}[h]
	\includegraphics[width=0.45\textwidth]{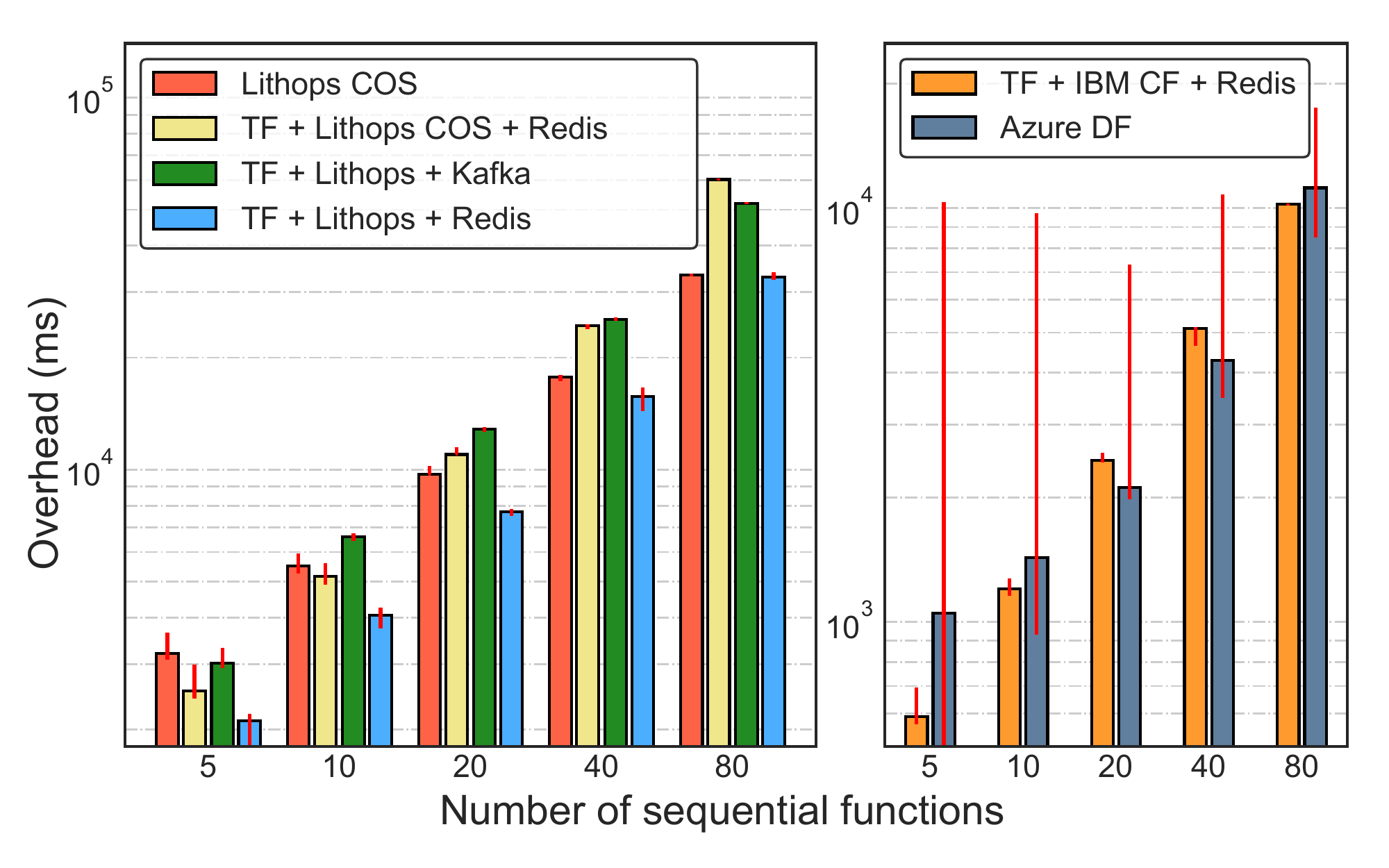}
	\centering
	\caption{Event sourcing overhead comparison for sequences. Lithops vs TF-Lithops on the left side. Triggerflow vs Azure Durable Functions on the right side.}
	\label{fig:pw_adf_sequence}
	\vspace{-8pt}
\end{figure}

\medskip
\noindent{\textbf{Sequential workflows.}} Figure \ref{fig:pw_adf_sequence} shows the overhead evolution when increasing the length of the sequence. The overhead added by both the native and external schedulers grows up linearly based on the number of functions in the sequence. As we can see, the results are very stable, meaning that the behavior is implementation-related, and not a problem with resources.

For the \emph{external scheduler}, we can see comparable performance between the original Lithops and our modified version for Triggerflow. Overhead evolves similarly in all scenarios. Lithops has to serialize and upload the function and the data to COS before executing it, creating overhead common for all scenarios. The remaining overhead comes from the place and the way these events are retrieved to recover the state of the execution (event sourcing). This means that the event source service---either COS, Kafka, or Redis---, has direct impact on these results.  For example, the main drawback of using COS in both the original (1) and Triggerflow (2) versions of Lithops is that they have to individually download the results from COS. This fact substantially increases the total time needed to execute a workflow, since for each step it has to retrieve all the previous events. In this case, for a workflow with $n$ steps, Lithops has to perform a total of $n(n+1)/2$ requests. In contrast, in the scenarios where Lithops does not use COS, and stores the events in a Kafka Topic (3) or a Redis Stream (4), it only needs one request to retrieve all the events in each step. Then, it only needs $n$ requests to these services to complete the execution of a workflow. If we compare scenarios 2 and 3, we see better performance if we use a Redis Stream instead of a Kafka Topic. This is mainly caused by the Kafka library, which adds a fixed overhead of $0.25$s each time the orchestration function is awaken and creates a consumer. This means that using a Kafka Topic as event store has a fixed overhead of $n*0.25$ seconds.

For the Triggerflow \emph{native scheduler}, it is important to note that the functions are already deployed in the cloud (in contrast with Lithops that has to serialize and upload them each time). Moreover, the orchestration code is execute within the TF-Worker that contains all the events loaded in memory, so it does not need to retrieve them from the event source (Kafka, Redis) in each step. Compared to ADF, we obtain similar overhead. As stated in the previous section, the overhead comes mainly from the fact that invoking an action in IBM CF service takes around $0.13$s. This means that, for a workflow of $n$ steps, Triggerflow has a fixed overhead of $n*0.13$ when using IBM CF.

\begin{figure}[h]
	\includegraphics[width=0.45\textwidth]{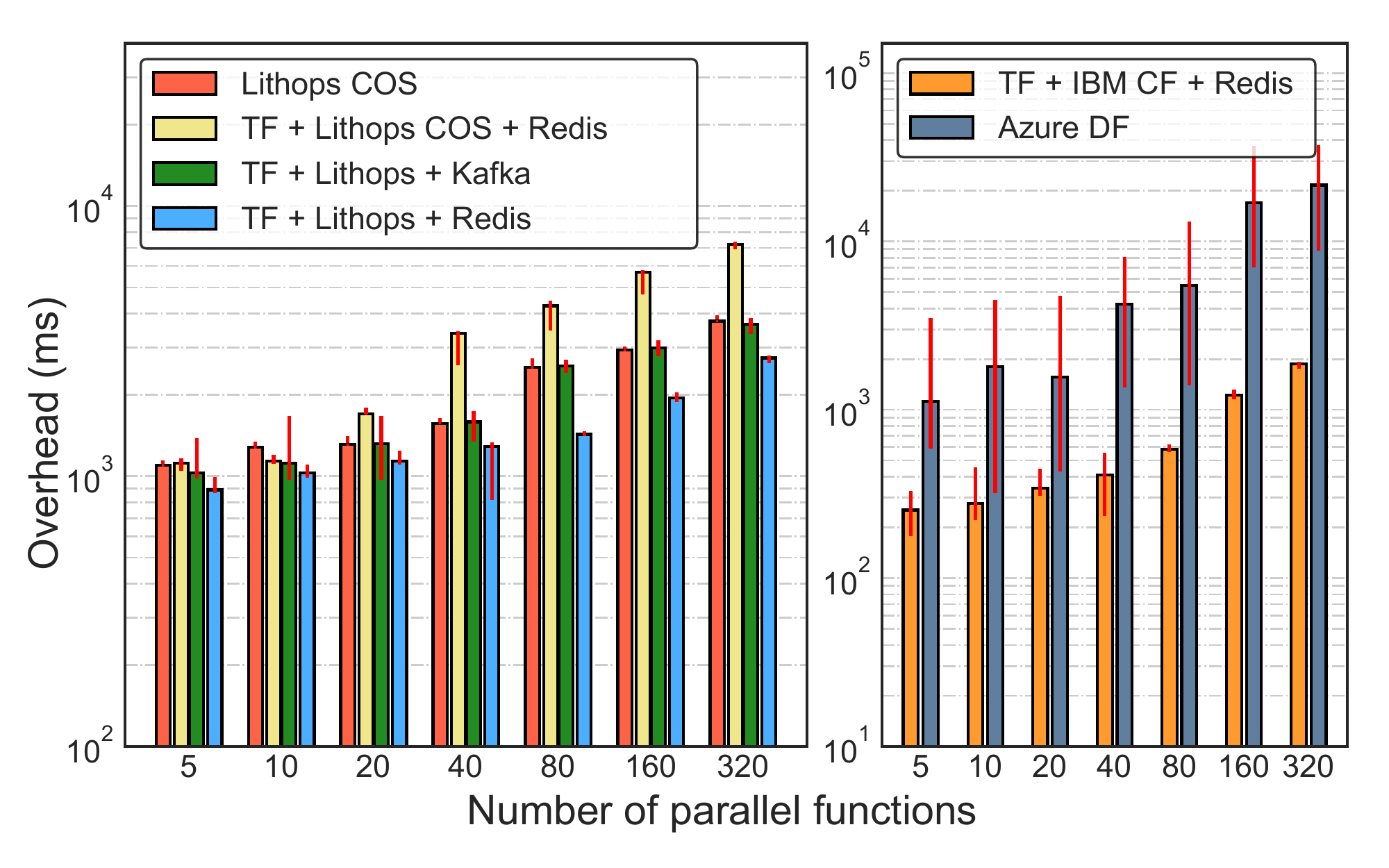}
	\centering
	\caption{Event sourcing overhead comparison for parallel workflows. Lithops vs TF-Lithops on the left side. Triggerflow vs Azure Durable Functions on the right side.}
	\label{fig:pw_adf_parallel}
	\vspace{-8pt}
\end{figure}

%\subsubsection{Parallel workloads}
\medskip
\noindent{\textbf{Parallel workflows.}} For this experiment, we evaluate the same scenarios described above. The results are depicted in Figure \ref{fig:pw_adf_parallel}. In this case, for the \emph{external scheduler}, the original Lithops and the Triggerflow Lithops version have also similar overhead, being scenario 4---which uses Redis as event store---the best approach. In the Kafka scenario (3), the overhead of $0.25$s described above is negligible, since in this experiment the orchestration function is awaken only once. The main difference in the performance between scenarios 1 and 2 is that the original Lithops is running all the time and polling the results as they are produced. In contrast, in the Triggerflow version of Lithops that uses COS (2), the TF-Worker first waits for all activations to finish to awake the orchestration function, that then has to retrieve all the events and results from COS. Finally, with the \emph{native scheduler}, Triggerflow is faster for parallel workflows compared to ADF. %Also, ADF presents unstable behavior. The variability can be explained with restrictions or difficulties to obtain resources past the $20$ concurrent functions. This makes us think that the increases in overhead are not strictly limited by the implementation, but by resource availability or provisioning in the Azure Cloud. In this service, we also see functions that are not scheduled until some have already finished.

\subsection{Scientific Workflows}

In this section, we will validate fault tolarence and feasibility for long running workflows of Triggerflow using real scientific workflows.

\subsubsection{Fault tolerance}

\begin{figure}[h]
	\includegraphics[width=0.45\textwidth]{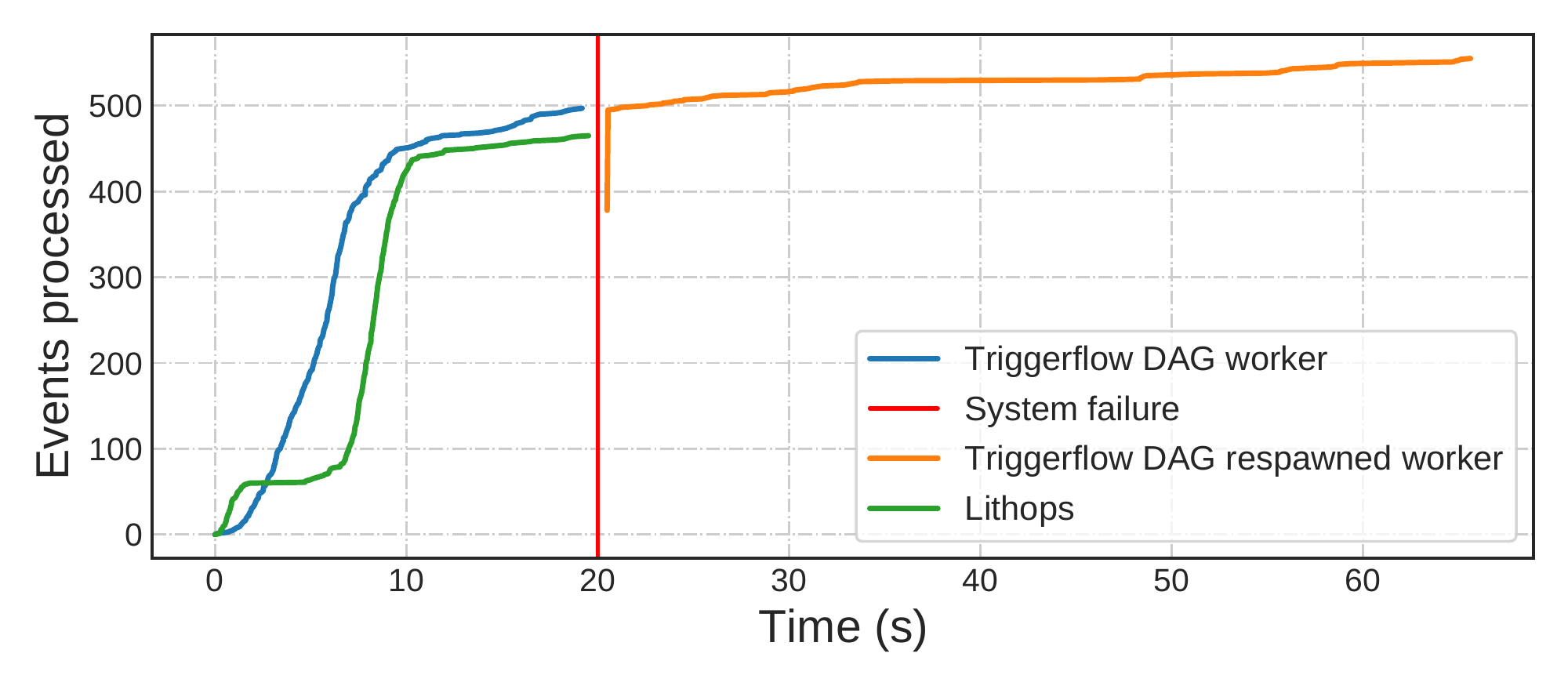}
	\centering
	\caption{Scientific workflow execution progression over time, with an intended system failure at the 20th second.}
	\label{fig:fault_tolerance}
	\vspace{-8pt}
\end{figure}

We adapted a geospatial scientific workflow, that was originally implemented with Lithops, to work with our DAGs interface. The objective of the workflow is to compute the evapotranspiration and water consumption of the crops from a set of partitioned geospatial data. Due to the nature of the workflow, and despite the optimizations applied, the workflow's execution time is similar to that provided by Lithops. The main difference lies in the workflow programming model: DAGs are more declarative and geared towards dissecting the workflow into independent tasks and their dependencies, while Lithops opts for a imperative map-reduce model. An important point in favor of Triggerflow is its automatic and transparent fault tolerance provided by the event source and trigger persistent storage. Figure \ref{fig:fault_tolerance} depicts the progression of a workflow run of the scientific workflow, using Kafka as the event source and Redis for the trigger storage. To check the system's fault tolerance, we intentionally stopped the execution of the Triggerflow worker and the Lithops execution in the 20th second of the workflow execution. Triggerflow rapidly recovers the trigger context from the database and the uncommitted events from the event source, and finishes its execution correctly. In contrast, Lithops stops and loses the state of the workflow, having to re-execute the entire workflow wasting time and resources.

\subsubsection{Long running workflows}

\begin{figure}[h]
	\includegraphics[width=0.3\textwidth]{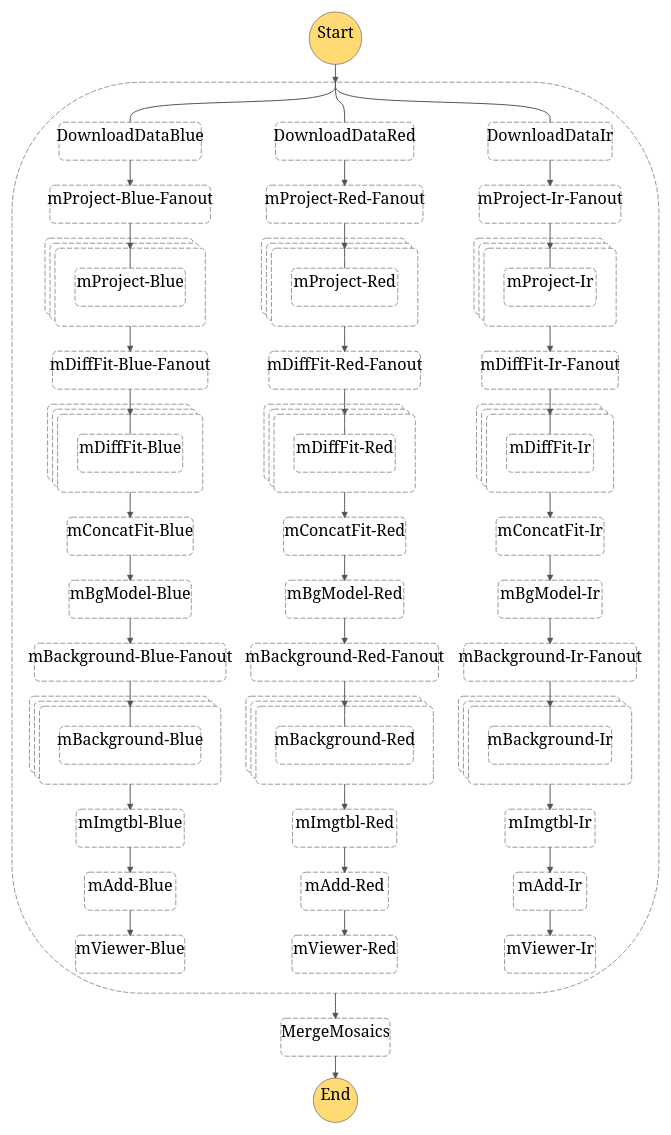}
	\centering
	\caption{Montage workflow represented as a Amazon Step Functions state machine.}
	\label{fig:montage_graph}
	\vspace{-8pt}
\end{figure}

We are now validating efficient resource utilization and auto-scaling to zero when orchestrating long-running scientific nested and parallel workflows represented as state machines. We want to run a long-running nested scientific workflow on Triggerflow and compare it to Amazon Step Functions. As a scientific workflow, we have implemented the classic Montage workflow, described in \cite{montage}. The Montage workflow is used to process astrological images and produce science-grade mosaics from multiple image data sets as if they were single images with a common coordinate system and projection. The Montage workflow consists of multiple consecutive steps that vary significantly in execution time, ranging from mere milliseconds up to minutes. Some of the steps can be executed as a parallel map, for example, the application of reprojection and background correction for every source image. Other steps aren't parallelizable and need to collect and combine data produced from a previous parallel step, like the calculation of parameters of the best-fit background model. At a higher level, we can produce an image for every RGB channel, in order to combine them at the end to produce a color image. The computation of these three images can also be run in parallel. In short, we have a nested workflow composed of three main parallel branches (one for each RGB channel), and that every branch executes the Montage workflow that has multiple consecutive steps, some of which can be mapped and run in parallel. A workflow diagram is presented in Figure \ref{fig:montage_graph}.

We have specified the Montage workflow using Amazon Step Language, since Montage workflow can be represented as a state machine with nested workflows, and we use Amazon Lambda to run the tasks. We have orchestrated the workflow on both Triggerflow and Amazon Step Functions. For this experiment, Triggerflow is deployed on Kubernetes with KEDA using Kafka as event stream.

\begin{figure}[h]
	\includegraphics[width=0.45\textwidth]{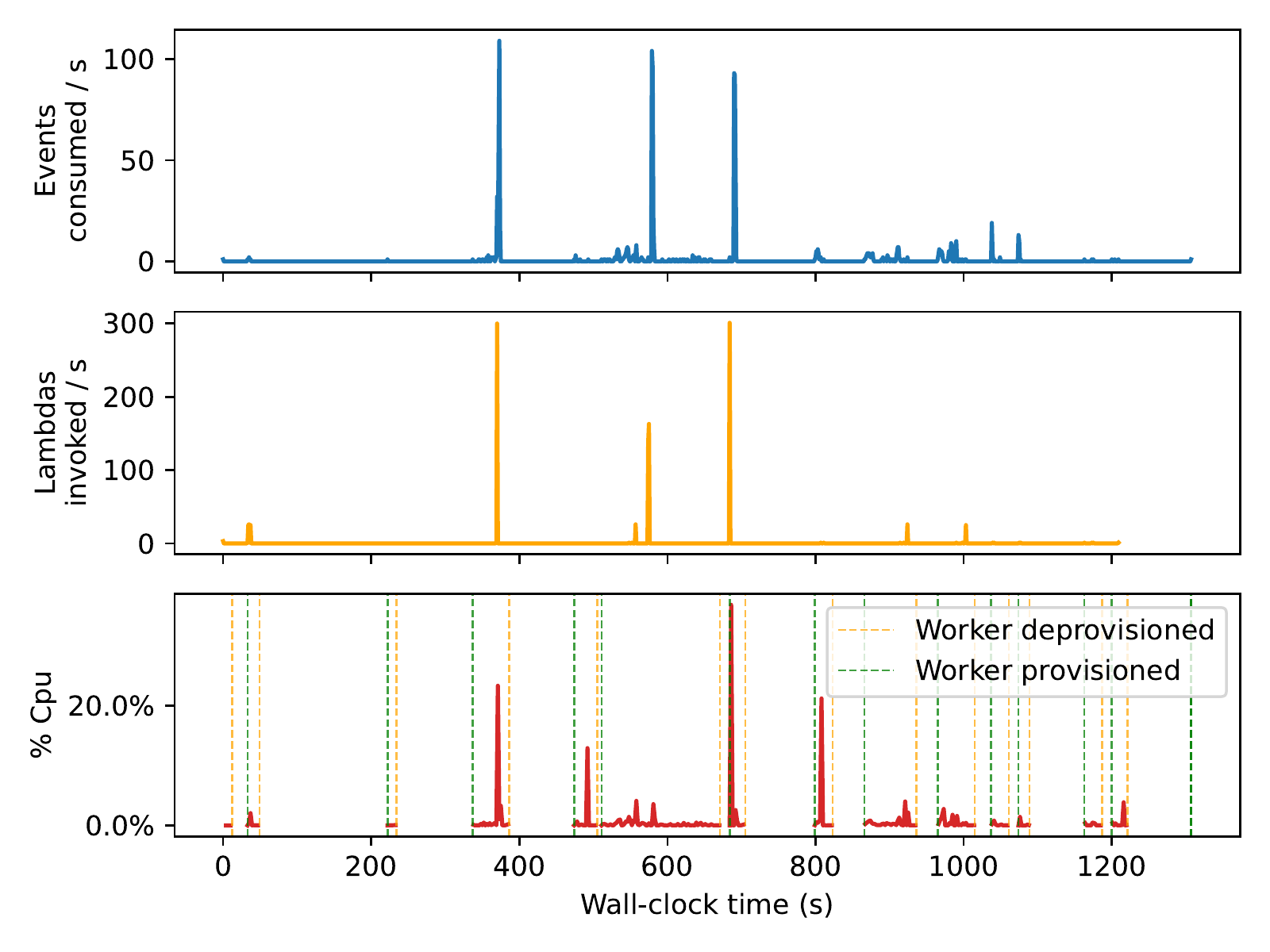}
	\centering
	\caption{(a) Events received per second, (b) functions invoked per second and (c) resources used in a Montage workflow execution using Triggerflow and KEDA.}
	\label{fig:montage_resources}
	\vspace{-8pt}
\end{figure}

\begin{figure}[h]
	\includegraphics[width=0.45\textwidth]{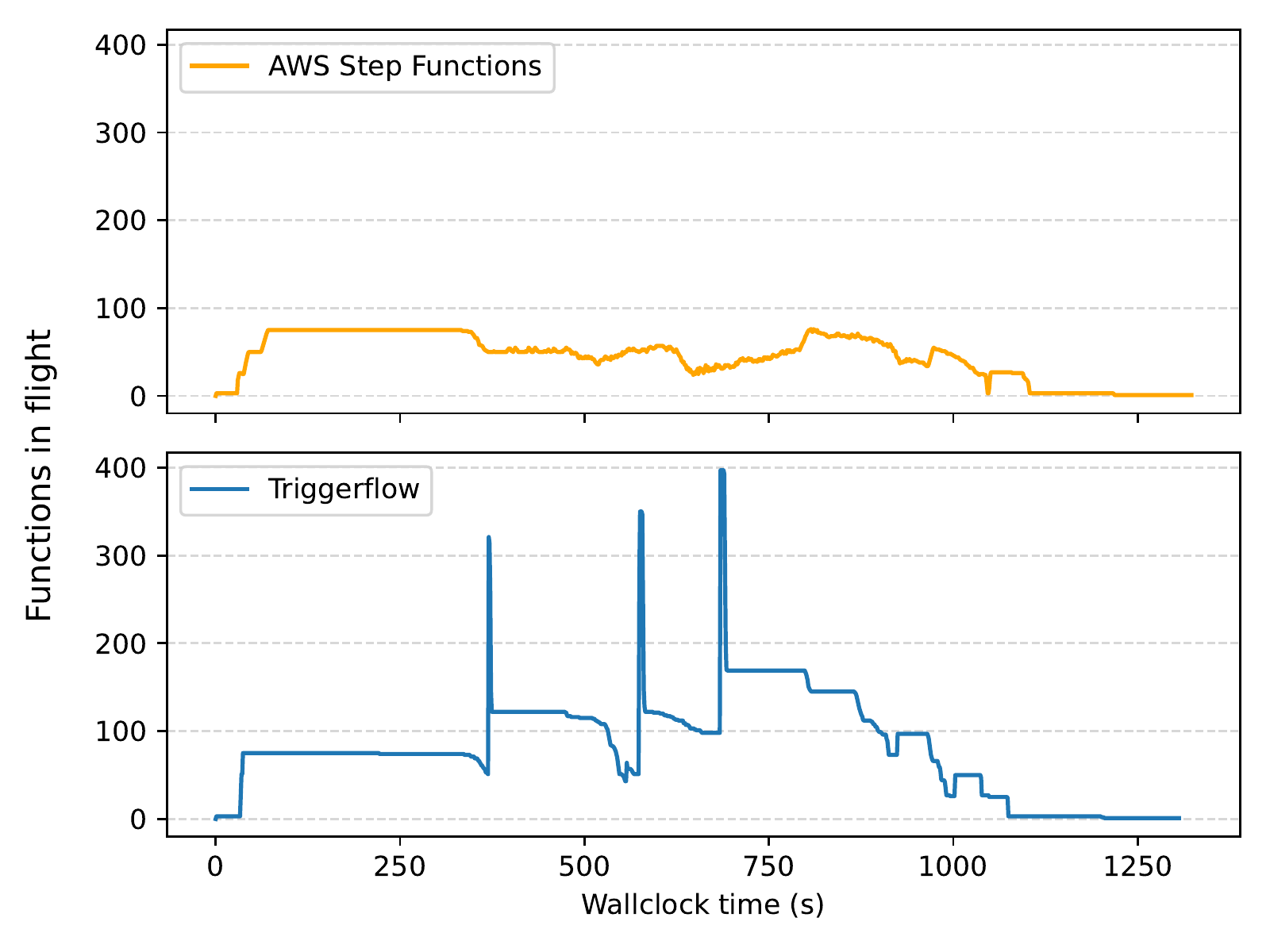}
	\centering
	\caption{(a) Total Lambda functions using Amazon Step Functions, (b) Total Lambda functions using Triggerflow}
	\label{fig:montage_histogram}
	\vspace{-8pt}
\end{figure}

Figure \ref{fig:montage_resources} represents a workflow execution on Triggerflow. We can see that KEDA automatically scales the worker pod down to zero while the long running tasks are being executed in Amazon Lambda. When the worker is up, it has 1 vCPU assigned. The worker is only executed when there is a state transition: KEDA only provisions the worker when there are events to be consumed from the broker. Every event is consumed, decoded into CloudEvents and processed through the trigger pipeline, executing the corresponding condition and action functions. At last, all events and trigger contexts are persisted in the storage database. The three peaks in events and invocations per second correspond to the most parallel task (mDiffFit) in the workflow. Finally, when a grace period of $10$ seconds passes without new events, KEDA scales down to zero the worker pod. We prove in this validation that Triggerflow makes an efficient use of system resources when orchestrating long running workflows.

Figure \ref{fig:montage_histogram} shows the total number of parallel functions being run in an execution. We can see that Triggerflow achieves a comparable execution time compared to Amazon Step Functions (Triggerflow is faster by approximately $30$ seconds). However, we achieve a greater level of function execution parallelism. This workflow could not be executed in Amazon Step Functions Express since it would exceed the permitted execution time of $5$ minutes.

%\subsubsection{Optimizations}
%
%To demonstrate Triggerflow's ability to introspect triggers with its Rich Trigger API, we have also implemented a service over the DAGs interface that automatically and transparently prewarms function containers on IBM Functions to increase the efficiency and overall parallelism, reduce total execution time and mitigate straggler functions effects in workflows that require high performance and throughput. Figure \ref{fig:prewarm} shows its effects. Thanks to Triggerflow's interception mechanisms we can also transparently apply other data pre-fetching optimizations in scientific workflows.

%\begin{figure}[t!]
	%\includegraphics[width=0.75\textwidth]{prewarm}
	%\centering
	%\caption{(a) Parallelism and total execution time in a sequential workflow that increases its map size and execution time every step. (b) Parallelism and total execution time of a single map task with high concurrency.}
	%\label{fig:prewarm}
	%\vspace{-8pt}
%\end{figure}

\subsection{Federated Learning orchestrator}

In this section, we will validate the Federated Learning orchestrator proposed in section \ref{sec:federated_learning}. The objective is to demonstrate how using Triggerflow to orchestrate a Federated Learning process, we can provide decoupling between the main server and the federated clients and failure flexibility.

We simulated a Federated Learning scenario using IBM Cloud Functions as federated clients and a process in a virtual server as the main server. To simulate the characteristic heterogeneity and proneness to failure of federated learning clients, we added a random factor that makes the function to take a random longer period of time and to randomly fail and never send a result. For the experiment, we used $50$ clients to train a model in three rounds and a result threshold of $65$\% .

\begin{figure}[h]
	\includegraphics[width=0.45\textwidth]{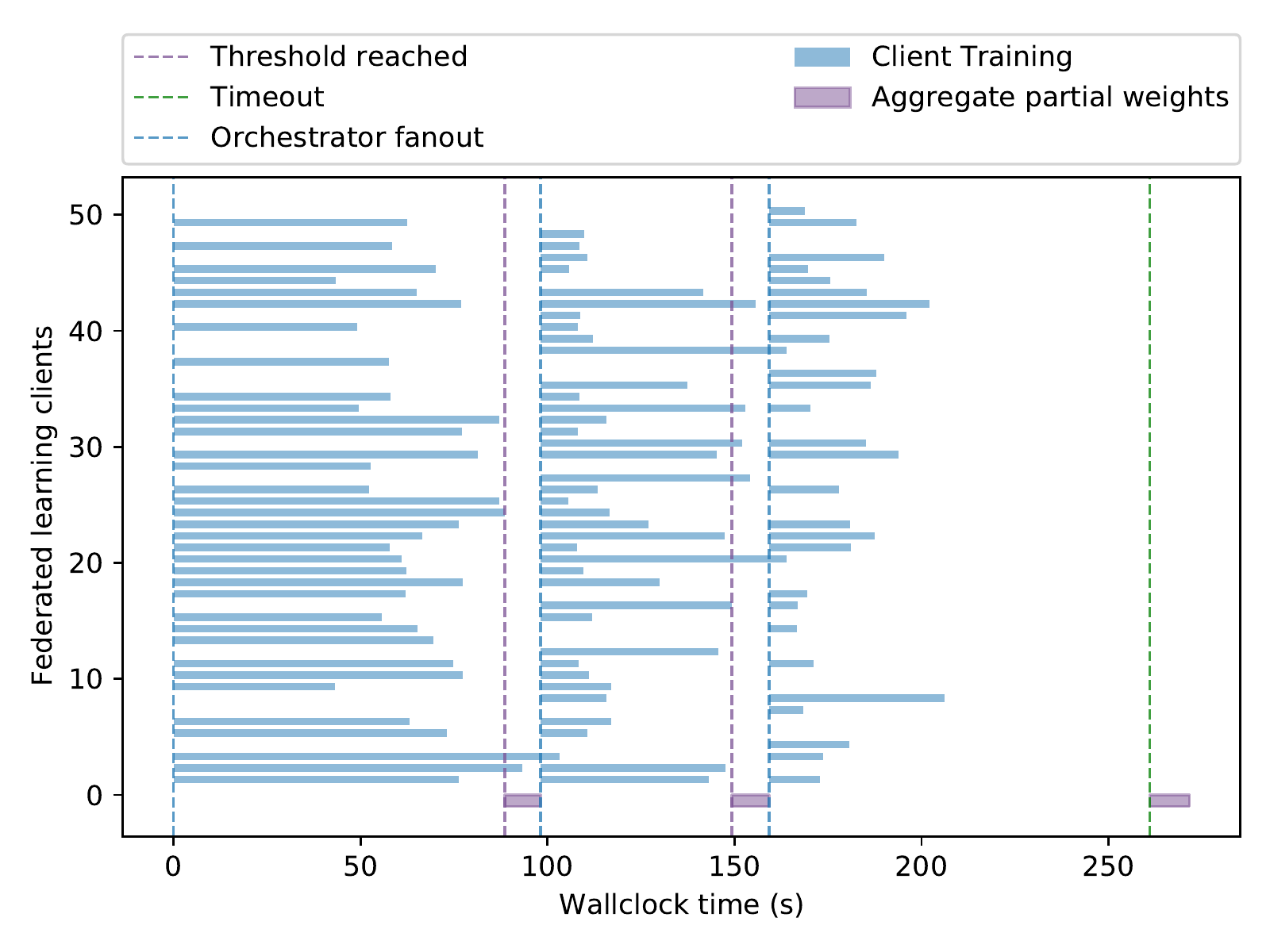}
	\centering
	\caption{A timeline of a Federated Learning process, using a pool of $50$ clients and three rounds.}
	\label{fig:federated_validation}
	\vspace{-8pt}
\end{figure}

Figure \ref{fig:federated_validation} represents the results of the federated learning process. For the first round, we can see that multiple clients participate in the training process and that each one takes a different amount of time to train the model. Some clients do respond to the invocation but will never send a result, thus simulating a network connection problem or other issues on the client side. Some other clients take a longer time than expected. These \textit{straggler} clients could slow down the whole process, this is why we set up a $65$\% threshold response. This means that the orchestrator will only wait for $65$\% of the total client pool to send their response (in this case, $32$ clients since we have a client pool size of $50$). When the threshold is reached, the \textit{aggregator} Trigger calls another IBM Cloud Function that recollects all results stored in IBM Cloud Object Storage and aggregates the partial model weights. Then, the \textit{aggregation function} fires the \textit{orchestrator trigger} through a termination event. The orchestrator trigger then invokes again the client pool to start another round. The second round passes similarly to the first round. However, in the third round, we can see that a lot of clients failed so they would never send the result. This could hang up the system. Despite that, a \textit{timeout event} is set up to prevent this case. This timeout sends an event to the \textit{aggregator trigger} to unblock the system so that the Trigger can take action on the failed round. In this case, it still aggregates the results and finishes the round successfully.

\subsection{Validation conclusions}

We have seen in this extensive validation section that our solution has met the proposed design goals.

We have used synthetic workloads to demonstrate the scalability, high performance and scale-to-zero serverless design of our architecture. We have also validated using real scientific workflows that event-based orchestration is suitable to provide fault tolerance and no performance loss for both long-running and short-running intense workflows.

Thanks to the flexibility provided by our programmable reactive actions, we have demonstrated that different workflow abstractions such as DAGs, State Machines or Workflow as Code can be orchestrated using events and triggers.

To finish off, we demonstrate that using generic triggers, we can build specialized event-based workflow abstractions like Federated Learning orchestrators. Using introspection mechanisms, we demonstrate that we can dynamically change the behavior of a workflow, for example by setting up a timeout event or by internally changing the state of a trigger.

\section{Conclusions}

In this article we have presented Triggerflow: a novel building block for controlling the life cycle of Cloud applications. As more applications are compiled to the Cloud, our system permits to encode their execution flow as reactive triggers in an extensible way. The novelty of our approach relies on four key aspects: serverless design, extensibility, support for heterogeneous workflows, and performance for high-volume workloads.

Triggerflow can become an extensible control plane for deploying reactive applications in the Cloud.  We implemented and validated different orchestration systems based on State Machines (ASF), Directed Acyclic Graphs (Airflow), Workflow as Code (Lithops), and a Federated Learning orchestrator. 

%we can implement over standard open source container technologies  We demonstrated that the system can support intensive event-based scientific workloads, and synchronize massively parallel jobs while providing low-latency choreography of serverless functions. %We are now implementing other schedulers like Federated Learning on top of Triggerflow.

As the number of event sources grows in many Cloud providers, trigger-based orchestration mechanisms will acquire more relevance in the future. In particular, the emergence of data-driven computations triggered by events \cite{gr2017whiz} is a good example of dynamic trigger-based orchestration.

Nevertheless, trigger-based approaches like Triggerflow still face serious challenges to become adopted. For instance, the observability of event-based flows is a complex open problem. Triggerflow has not addressed the problem of inferring the structure of a workflow from a set of events. We only provide reactive actions to concrete events and generate triggers from pre-defined workflows. In addition, debuggability and developer experience are very important to enable the adoption of such event-based models. As an open source project, Triggerflow would clearly benefit from tools and user interfaces to simplify the overall observability and life-cycle support of the system. 

%flow is not targeting now business workflows, its extensibility permits to support them in the future. In this case, business domains typically do not require low latency or supporting high parallel jobs, but instead they need stronger system reliability and manageability with auditing and compliance requirements.

%Business workflows that are designed for business use cases and business domains typically do not require low latency and supporting high parallel jobs may be less impoerant than system realibility and mangeability with auditing and complaince requirements. Using Triggerflow extensiility it should be possible to support business workflows, even though Triggerflow is not targeting business domain currently.

%We implemented Triggerflow on top of standard open source technologies like CNCF Kubernetes, Knative (Serving and Eventing), and CloudEvents. This approach makes our proposal reproducible and portable to different Cloud providers and on-premise environments. 

%In summary, we are offering an extensible control plane for deploying reactive applications in the Cloud. We foresee more applications and frameworks leveraging these extensible control plane to optimize their execution in Cloud settings.

%Furthermore, many of the abstractions available in Triggerflow (dynamic triggers, custom filters, termination events, shared context) may be adopted by the event router.

\section*{Acknowledgments}
%% CloudButton = 825184
This work has been partially supported by the EU Horizon 2020 programme under grant agreement No 825184.

\bibliography{article}

\end{document}